\documentclass{emulateapj}

\slugcomment{Accepted for Publication in The Astrophysical Journal}

\shorttitle{Large Area Survey for $z=7$ Galaxies}
\shortauthors{Ouchi et al.}

\begin{document}

\title{Large Area Survey for $\lowercase{z}=7$ Galaxies
in SDF and GOODS-N:\\
Implications for Galaxy Formation and Cosmic Reionization
\altaffilmark{\ddag}
}

\author{
Masami Ouchi        \altaffilmark{1,2},
Bahram Mobasher     \altaffilmark{3},
Kazuhiro Shimasaku  \altaffilmark{4,5},
Henry C. Ferguson   \altaffilmark{6,7},\\
S. Michael Fall           \altaffilmark{6},
Yoshiaki Ono        \altaffilmark{4},
Nobunari Kashikawa  \altaffilmark{8},
Tomoki Morokuma     \altaffilmark{8,9},\\
Kimihiko Nakajima   \altaffilmark{4},
Sadanori Okamura    \altaffilmark{4,5},
Mark Dickinson      \altaffilmark{10},\\
Mauro Giavalisco    \altaffilmark{11},
Kouji Ohta          \altaffilmark{12}
}

\altaffiltext{1}{Observatories of the Carnegie Institution of Washington,
        813 Santa Barbara St., Pasadena, CA 91101}
\altaffiltext{2}{Carnegie Fellow; ouchi \_at\_ obs.carnegiescience.edu}
\altaffiltext{3}{Department of Physics and Astronomy, University of California, Riverside, CA 92521, USA}
\altaffiltext{4}{Department of Astronomy, School of Science, University of Tokyo, Tokyo 113-0033, Japan}
\altaffiltext{5}{Research Center for the Early Universe, School of Science,
University of Tokyo, Tokyo 113-0033, Japan}
\altaffiltext{6}{Space Telescope Science Institute, 3700 San Martin Drive, Baltimore, MD 21218}
\altaffiltext{7}{Department of Physics and Astronomy, Johns Hopkins University, 3400 N. Charles St., Baltimore, MD 21218}
\altaffiltext{8}{Optical and Infrared Astronomy Division, National Astronomical Observatory, Mitaka, Tokyo 181-8588, Japan}
\altaffiltext{9}{Research Fellow of the Japan Society for the Promotion of Science}
\altaffiltext{10}{NOAO, 950 N. Cherry Avenue, Tucson, AZ 85719}
\altaffiltext{11}{Department of Astronomy, University of Massachusetts, Amherst, MA 01003}
\altaffiltext{12}{Department of Astronomy, Kyoto University, Kyoto 606-8502, Japan}

\altaffiltext{\ddag}{Based on data obtained with 
the Subaru Telescope, the NASA/ESA Hubble Space Telescope (HST),
and Spitzer Space Telescope. The Subaru Telescope is 
operated by the National Astronomical Observatory of Japan.
HST is operated by the Association of Universities for
Research in Astronomy (AURA), Inc., under NASA contract NAS5-26555.
The Spitzer Space Telescope is operated by the Jet Propulsion 
Laboratory, California Institute of Technology 
under a contract with NASA.
}

\begin{abstract}
We present results of our large-area survey for
$z'$-band dropout galaxies at $z=7$ 
in a 1568 arcmin$^2$ sky area covering the SDF
and GOODS-N fields. Combining our ultra-deep 
Subaru/Suprime-Cam $z'$- and $y$-band
($\lambda_{\rm eff}=1\mu{m}$) images with
legacy data of Subaru and HST, 
we have identified 22 bright $z$-dropout galaxies
down to $y=26$, one of which has 
a spectroscopic redshift of $z=6.96$ 
determined from Ly$\alpha$ emission.
The $z=7$ luminosity function (LF)
yields the best-fit Schechter parameters of
$\phi^*=0.69_{-0.55}^{+2.62} \times 10^{-3}$Mpc$^{-3}$,
$M_{\rm UV}^* = -20.10\pm 0.76$ mag, and
$\alpha = -1.72\pm 0.65$, 
and indicates a decrease from $z=6$
at a $>95$\% confidence level.
This decrease is beyond the cosmic variance 
in our two fields, which is estimated 
to be a factor of $\lesssim 2$. 
We have found that the cosmic star formation rate density 
drops from the peak at $z=2-3$ to $z=7$ roughly by a factor of
$\sim 10$ but not larger than $\sim 100$.
A comparison with the reionization models
suggests either that 
the Universe could not be totally ionized by only galaxies
at $z=7$, or more likely that properties of
galaxies at $z=7$ are different from those at low redshifts
having, e.g., a larger escape fraction ($\gtrsim 0.2$), 
a lower metallicity, and/or a flatter IMF.
Our SDF $z$-dropout galaxies
appear to form $60$-Mpc long filamentary structures,
and the $z=6.96$ galaxy with Ly$\alpha$ emission
is located at the center of an overdense region
consisting of four UV bright dropout candidates,
which might
suggest an existence of a well-developed ionized bubble
at $z=7$.
\end{abstract}


\section{Introduction}
\label{sec:introduction}

Recent results from deep galaxy surveys have raised exciting
questions about cosmic reionization and the early phases of galaxy formation. 
These surveys have extended the redshift frontier to $z\sim 6-10$, 
using various techniques \citep{iye2006,stark2007,ota2008,richard2008,
bouwens2008,bouwens2009a,bouwens2009b,oesch2009a,oesch2009b,
castellano2009,mclure2009a,mclure2009b,bunker2009,hickey2009,
henry2007,henry2008,henry2009,bradley2008,zheng2009,sobral2009}. 
These observations have reached near the epoch of reionization
at $z>6$, which is suggested by \citet{fan2006}
who find that the Gunn-Peterson (GP) optical depths of SDSS QSOs 
significantly increase at $z\sim 6$.
Although this increase can be explained 
by the extrapolation from low $z$ to $z>6$ 
via adoption of log-normal distribution 
for optical depths \citep{becker2007},
it is also true that GP optical depths appear 
to increase significantly at around $z=6$.
The ionizing sources at $z\simeq 6$ are 
efficiently identified by $i'$-dropout technique 
(\citealt{stiavelli2004}; cf. \citealt{bunker2004}).
However, the relation between reionization and
ionizing sources, i.e. galaxies, are still not clear 
in our understanding of cosmic reionization.
Since the WMAP5 polarization data indicate
possible scenarios of an instantaneous reionization 
at $z=11\pm 1.4$ and an extended reionization 
at $z\sim 6-11$ \citep{dunkley2009}, 
the ionizing photon production rate
of galaxies at this epoch would constrain
these models of reionization history.
It is suggested that a substantial fraction of galaxies have completed
their starburst phase before $z\simeq 6$. \citet{eyles2007} have shown that
about 40\% of the IRAC detected $z\simeq 6$ dropout galaxies exhibit 
a significant Balmer break, indicating that these are post-starburst galaxies 
whose major formation phase ended at $z>6$. 
\footnote{
Recently, \citet{schaerer2009} argue that
\citet{eyles2007} would overestimate the Balmer breaks
by the contributions of nebular emission to their broad band photometry.
}
It is important to understand when and how
the progenitors of these post-starburst galaxies were formed. To address
these questions, we need to study $z\gtrsim 7$ galaxies over a wide range of
magnitudes to derive UV luminosity function (LF) and density.

Currently, only a handful of $z\simeq 7$ dropout galaxy candidates are 
photometrically identified, due to observational difficulties of identifying
$z\simeq 7$ dropout galaxies whose detectable continuum is 
redshifted to $1\mu$m (e.g. \citealt{bouwens2008,oesch2009a}).
Although Hubble Space Telescope (HST) 
images can reach as deep as $\sim 29$ AB magnitude in near-infrared (NIR) bands 
with the state-of-the-art Wide Field Camera 3 (WFC3),
the number of $z\simeq 7$ dropout galaxy candidates is still limited up to
$\simeq 10-20$, so far (\citealt{oesch2009b,mclure2009b,bunker2009}; see
also \citealt{bouwens2009b}).
Moreover, the present studies cover only small
areas ($\simeq 5$ arcmin$^2$ for the HST/WFC3 studies and
$\simeq 100$ arcmin$^2$ even for recent VLT/HAWK-I observations; 
\citealt{castellano2009,hickey2009})
or small volumes by the gravitational lensing technique
(e.g. $\sim 100$ Mpc$^3$; \citealt{stark2007}).
Although gravitational lensing technique can probe
very faint high-$z$ galaxies that cannot be found 
in blank fields with the current facilities
\citep{stark2007,bradley2008,bouwens2009a,zheng2009}, 
these pencil beam surveys suffer from large uncertainties due to 
cosmic variance, and miss a population of 
bright galaxies at $z>7$, which plays an important role
in determining UV luminosity density for cosmic reionization 
and in understanding forming massive galaxies in their
major star-formation phase.

Motivated by this, we have conducted
a wide-field $z$-dropout galaxy survey 
by performing deep $z'$ and $y$-band 
imaging down to $y=26.0$ with 
Subaru/Suprime-Cam \citep{miyazaki2002}, exploiting 
the excellent red sensitivity of the wide-field
optical camera mounted on the 8.3m Subaru telescope.
Our program realizes a very large volume survey
for $z$-dropout galaxies with 
the reasonably deep magnitude limit,
which strongly constrains the bright-end of UV LF at $z=7$.
In this paper, we present our bright
$z$-dropout galaxy candidates found by 
our Subaru observations, and constrain cosmic
star-formation history and reionization in conjunction
with faint $z$-dropout galaxies identified by 
the recent deep HST/WFC3 studies.
We describe our observations and 
$z=7$ $z$-dropout galaxy samples
in \S \ref{sec:data} and \S \ref{sec:samples}, respectively.
We show the UV LF and UV luminosity
density of $z=7$ galaxies 
in \S \ref{sec:luminosity_function}.
We discuss the cosmic star-formation history,
cosmic reionization, and 
the distribution of our $z$-dropout galaxies 
in \S \ref{sec:discussion},
and summarize our results in \S \ref{sec:conclusions}.
Throughout this paper, magnitudes are in the AB system.
We adopt
$(h,\Omega_m,\Omega_\Lambda, \Omega_b h^2, n_s,\sigma_8)=
(0.7,0.3,0.7,0.02,1.0,0.9)$.

\begin{figure}
\epsscale{1.2}
\plotone{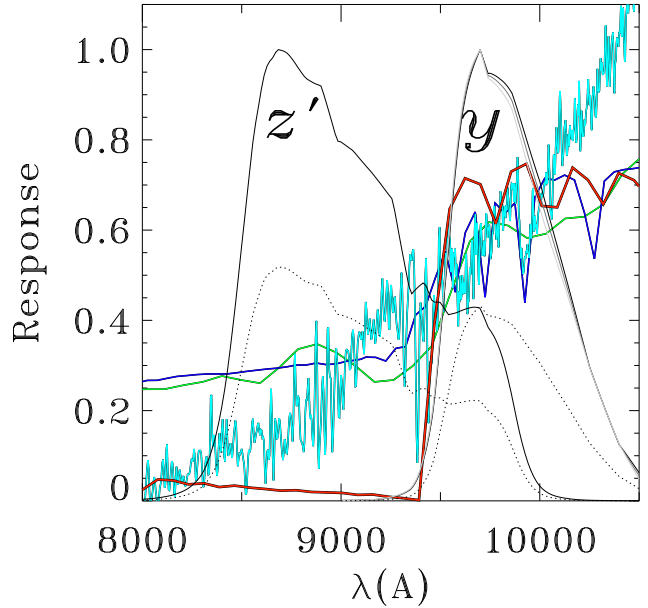}
\caption{
Response curves of $y$ and $z'$ bands,
together with model spectra of a $z\sim 7$ galaxy
and foreground interlopers.
Black solid lines represent the response curves
whose peaks are normalized to $1.0$, while the
black dotted lines denote those showing
the total system throughput. 
Dark and light gray lines are the
normalized total $y$-band response curves 
for the flux-weighted SDF and GOODS data.
All of these response curves include 
the CCD quantum efficiency, 
transmission+reflection of telescope+instrument optics, 
and atmospheric absorption.
Red line is the model spectrum mimicking
the one of a typical dropout galaxy \citep{papovich2001},
but redshifted to $z=6.73$. 
Green line is a template spectrum
of the typical local elliptical (old) galaxy 
placed at $z=1.4$ \citep{coleman1980}.
Because the age of the Universe at $z=1.4$ is only
$4-5$ Gyr which is younger than 
the typical age of local ellipticals,
the oldest galaxies at $z=1.4$ are probably younger than 
the local elliptical of this template. Thus,
even the oldest galaxies at $z=1.4$ should
have a bluer SED than this elliptical template.
Blue line plots a dusty-starburst spectrum at $z=1.5$,
which is a reproduction of the typical dusty-starburst SED \citep{cimatti2002}.
Cyan line shows a Galactic T3 dwarf star, one of the reddest
Galactic dwarf stars \citep{knapp2004}.
\label{fig:plot_sed_paper}}
\end{figure}

\section{Data}
\label{sec:data}

\subsection{Observations and Data Reduction}
\label{sec:observations_and_data_reduction}

We carried out deep $y$- and $z'$- band imaging survey 
with Suprime-Cam in the Subaru Deep Field 
(SDF; \citealt{kashikawa2004})
and GOODS-N \citep{giavalisco2004a} during
the last three years from 2006 to 2009.
The $y$ band is a custom broad-band filter centered at $1\mu$m.
This filter was originally made for identifying
$z\simeq 6$ galaxies by the two-color Lyman break
selection technique \citep{shimasaku2005}.
Although this filter was formerly called
$z_R$ filter in \citet{shimasaku2005},
we refer to it as $y$ filter for sake of simplicity.
Figure \ref{fig:plot_sed_paper} shows the total response of
the $y$ band filter (central wavelength of 9860\AA\ and FWHM of 590\AA), 
which includes the CCD quantum efficiency, transmission+reflection of 
telescope+instrument optics, and atmospheric absorption.
Each of SDF and GOODS-N field is covered by
one pointing of Suprime-Cam whose FoV is 918 arcmin$^2$.
Table \ref{tab:obs} summarizes our observations and the $y$ and $z'$ 
data that we used.
Before 2007, we took $y$ data with MIT-Lincoln Laboratory (MIT-LL)
CCDs installed in Suprime-Cam \citep{miyazaki2002}, while 2009 data 
were acquired with the new fully-depleted Hamamatsu CCDs \citep{kamata2008}.
In our analysis, we also use archival $y$ and $z'$
images taken in 2001-2006 to make the deepest
stacking of $y$ and $z'$ data available in SDF and GOODS-N.
The total integration time of our $y$ band images are
26 and 33 hours in SDF and GOODS-N, respectively.
The $z'$ image of SDF is produced via a number of variable
object studies such as supernovae \citep{poznanski2007}, active galactic
nuclei (Morokuma T. et al. in preparation), and high proper motion stars 
\citep{richmond2009}.

Our data were reduced using Suprime-cam Deep Field REDuction
package (SDFRED; \citealt{yagi2002,ouchi2004}).
We have found that the total throughput in the $y$ band is improved
by nearly a factor of 2 in 2009 with the new Hamamatsu CCDs,
but that the shapes of $y$-band total response curves
are almost identical between the data of MIT-LL (taken before 2007) 
and Hamamatsu CCDs (taken in 2009; Figure \ref{fig:plot_sed_paper}).
Thus, we combine these $y$-band data taken with MIT-LL and Hamamatsu CCDs.
Figure \ref{fig:plot_sed_paper} displays the normalized total $y$-band response curves
of
the Hamamatsu CCDs and 
the flux-weighted MIT-LL+Hamamatsu CCDs for the SDF and GOODS data.
The differences between all the response curves are negligibly small.
Since we find a very small difference of $\lesssim 0.02$ mag between the response curves of
flux-weighted SDF and GOODS-N even with the model spectra of extremely red objects such as
L/T-dwarf stars and $z$-dropout galaxies (\S \ref{sec:luminosity_function}),
we only use the $y$-band response curve of the flux-weighted SDF in our analysis.
The $3\sigma$ sky noise of the reduced images
are $(y,z') = (26.4, 27.7)$ and $(26.2, 26.9)$ magnitudes
in a $1''.8$-diameter circular aperture
for SDF and GOODS-N, respectively.

We mask contaminated areas with halos of bright stars 
and CCD blooming or low signal-to-noise (S/N) regions
caused by dithering at around the edge of the FoV.
The sky area after the masking is 810 and 758 arcmin$^2$
in SDF and GOODS-N, respectively. The total area is 1568 arcmin$^2$.
The positions of $y$ and $z'$ are aligned based on hundreds of
stellar objects commonly detected in both images. 
After the PSF sizes of these images are matched,
FWHM sizes of PSFs are estimated to be 
$\simeq 0''.91$ and $\simeq 0''.87$
in SDF and GOODS-N, respectively.

\subsection{Matched Images}
\label{sec:matched_images}

Except for the $y$ and $z'$ bands, we use the legacy 
ground-based deep optical imaging data for SDF \citep{kashikawa2004} 
and GOODS-N \citep{capak2004}, and the deep HST/ACS v2.0 and 
Spitzer/IRAC v0.3 available for 
the central $\simeq 160$ arcmin$^2$ field of GOODS-N
\citep{giavalisco2004a,dickinson2003}.
\footnote{
Note that HST/ACS v2.0 data
are significantly deeper than the previously released images.
}
These data are registered with stellar objects 
to match the coordinates of our Suprime-Cam $y$ and $z'$ images 
in SDF and GOODS-N.
Table \ref{tab:limitingmag} summarizes all of imaging data used in our study.
Again, the PSF sizes of our SDF $y$ and $z'$ images
are matched to that of the public SDF images with a PSF FWHM of $0''.99$. 
Note that the public ground-based GOODS-N data include
$U$, $V$, and $R$-band images with a relatively poor seeing 
($FWHM\simeq 1''.5$). We cannot homogenize the seeing sizes
of GOODS-N images without a significant loss of S/N ratios of 
our $y$ and $z'$ images.
Because our purpose of the $UBVR$ photometry is not to measure a color 
defined by the same PSF+aperture but to confirm non-detections,
we place the upper limits of detection with a large, $3''$-diameter, aperture size
for those $U$, $V$, and $R$ images.

During the observations, we took images of spectrophotometric 
standard star of G191-B2B with $y$ band filter in GOODS-N
\citep{oke1990,bohlin1995}.
The standard star was observed $4$ times
under photometric condition.
We calculate photometric zero-points 
from the standard star data.
The photometric zero points of the other images, i.e.
GOODS-N $z'$, SDF $y$, and SDF $z'$, are determined
by matching the zero points with those of images
taken by \citet{capak2004},
\citet{shimasaku2005}, and \citet{kashikawa2004}, 
respectively.
We check these photometric zero points 
based on colors of stellar objects in our field
and 175 Galactic stars calculated from
spectra given in \citet{gunn1983}. 
We find that the colors of stellar objects in our data
are consistent with those of \citeauthor{gunn1983}'s 
(\citeyear{gunn1983}) stars
within $\simeq 0.03$ magnitude.

\section{Catalogs and Samples}
\label{sec:samples}

\subsection{Photometric Catalogs}
\label{sec:photometriccatalogs}

Source detection and photometry are performed using
SExtractor 
\citep{bertin1996}.
The $y$ images are chosen for our source detection.
We measure $1''.8$-diameter aperture magnitudes,
and define a $z'-y$ color with these aperture magnitudes
obtained by the dual image mode of
SExtractor.
We correct the magnitudes of objects
for Galactic extinction of 
$E(B-V)=0.018$ (SDF) and
$0.012$ (GOODS-N; \citealt{schlegel1998}).
A total of 63,740 and 55,559 objects
are identified down to the $4\sigma$ detection limits
in SDF ($y=26.1$) and GOODS-N ($y=25.9$), respectively.

\begin{figure}
\epsscale{1.2}
\plotone{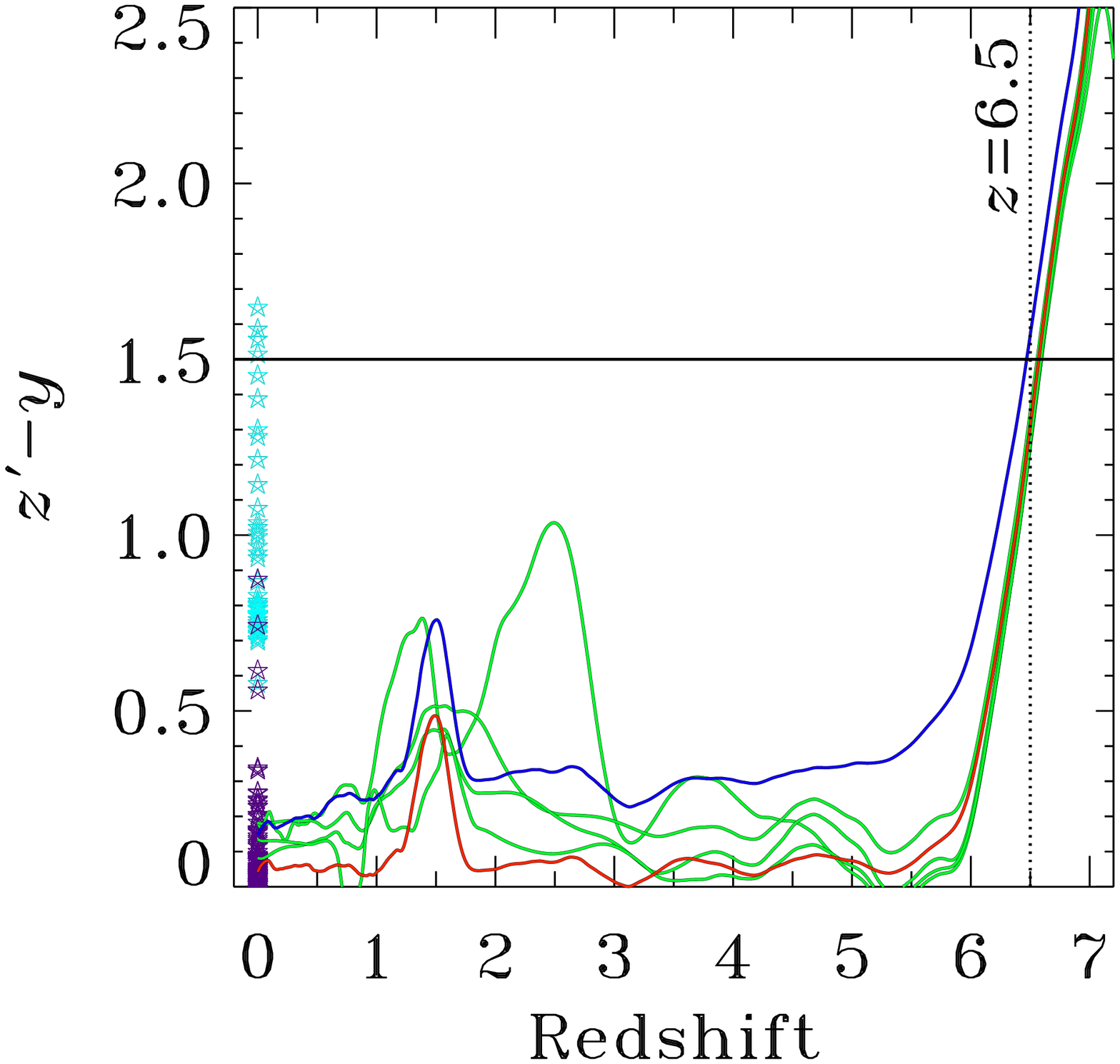}
\caption{
Colors of $z'-y$ for various objects
as a function of redshift. Red and blue lines represent 
the models of dropout and dusty-starburst galaxies
that are reproductions of the typical SEDs shown 
in \citet{papovich2001} and \citet{cimatti2002}, respectively.
Green lines indicate elliptical, Sbc, Scd, and irregular galaxies 
\citep{coleman1980}. Purple and cyan star marks are Galactic stars
\citep{gunn1983} and L1-L9/T0-T9 dwarf stars \citep{knapp2004}.
Black solid line indicates our color selection criterion, $z'-y>1.5$.
The black dotted line marks a redshift, $z=6.5$,
that is roughly a lower limit of our selection.
\label{fig:redshift_color_paper}}
\end{figure}

\subsection{Photometric Samples}
\label{sec:photometricsamples}

We isolate $z$-dropout galaxy candidates at $z\sim 7$ 
from foreground interlopers with our photometric data. 
Figure \ref{fig:plot_sed_paper} plots 
the model spectra of a $z\sim 7$ galaxy as well as 
red objects including low-$z$ galaxies and
one of the reddest T dwarf stars \citep{knapp2004}
which are likely to be prominent interlopers in our photometric sample.
Figure \ref{fig:plot_sed_paper} shows that the spectral feature of 
a significant 1216\AA\ trough for $z\sim 7$ galaxies is
clearly different from the spectral shapes of foreground 
objects at the $z'$- and $y$-band wavelengths.

In Figure \ref{fig:redshift_color_paper}, we present
predicted $z'-y$ colors as a function of redshift
for these model spectra. This figure demonstrates that
no objects except $z\sim 7$ galaxies and some late-type dwarf stars 
have colors redder than $z'-y\gtrsim 1.5$.
Since the wavelength coverages of $z'$- and $y$-bands are very close
(Figure \ref{fig:plot_sed_paper}), the Ly$\alpha$ trough of $z\sim 7$ galaxies
can be more clearly distinguished from the red continuum of
foreground objects. Indeed, the elliptical galaxy of \citet{coleman1980} 
is bluer than $z\gtrsim 6.5$ galaxies by $\Delta(z'-y)\gtrsim 0.4-0.5$ 
even at its reddest color. A strong $z'-y$ color cut can isolate $z\sim 7$ galaxies
with a small fraction of foreground interlopers.
On the other hand, the recent studies of $z\sim 7$ galaxies 
apply a relatively weak color criterion for the Ly$\alpha$ trough, 
e.g. $z'-Y_{105} \gtrsim 1$ or $z'-J_{110} \gtrsim 1$,
with a broad band of $Y$/$J$ whose central wavelength is redder than 
that of our $y$ band
\citep{bouwens2008,oesch2009b,bunker2009,castellano2009,hickey2009}.
The colors of $z'-Y_{105} \sim 1$ and $z'-J_{110} \sim 1$ fall below 
the red peaks of the \citeauthor{coleman1980}'s (\citeyear{coleman1980}) 
elliptical galaxy (see, e.g., Figure 2 of \citealt{bunker2009}),
and these studies cannot remove such foreground red galaxies
with a $z'-Y_{105}$ or $z'-J_{110}$ color alone. Instead,
these studies can distinguish foreground red galaxies
with their available deep NIR ($J$, $H$, and/or $K$) photometry 
(e.g. Figure 3 of \citealt{bunker2009}). The idea of our
candidate selection is to discriminate foreground red galaxies
without NIR photometry but with a color cut stronger
than the color criterion of the other studies.
In addition to a Ly$\alpha$ trough, 
a rest-frame far UV continuum below Lyman break ($912$\AA) is damped
by IGM absorption with a large Lyman continuum opacity
\citep{inoue2008}. Because no such continuum should be identified 
at a wavelength shorter than $7000(\simeq 912\times [1+6.7])$\AA\ 
for sources at $z\gtrsim 6.7$, non-detection criteria
should be given in $U$, $B$, $V$, and $R$ bands whose
bandpasses are bluer than $7000$\AA.
From the model colors and the lack of a far UV continuum,
we define the selection criteria of $z\sim 7$ galaxies:
{\footnotesize
\begin{eqnarray}
\label{eq:zdropselection}
z'-y>1.5\ \ \ \&\ \ \ U>U_{2\sigma}\ \ \ \&
\ \ \ B>B_{2\sigma}\ \ \ \&\ \ \ V>V_{2\sigma}\ \ \ \&
\ \ \ R>R_{2\sigma}
\end{eqnarray}
}
where
$U_{2\sigma}$, $B_{2\sigma}$, $V_{2\sigma}$, and $R_{2\sigma}$ are
the $2\sigma$ limiting magnitudes of $U$, $B$, $V$, and $R$ images,
respectively. The $U$-band criteria is only applied to 
the objects in GOODS-N, since there are no public $U$-band data
in SDF. The $2\sigma$ limiting magnitudes are
$(B_{2\sigma}, V_{2\sigma}, R_{2\sigma}) = 
(29.4, 28.6, 28.8)$ in SDF and
$(U_{2\sigma}, B_{2\sigma}, V_{2\sigma}, R_{2\sigma}) = 
(27.3, 28.0, 26.7, 27.0)$ in GOODS-N.
We select $z\sim 7$ galaxy candidates
with these photometric criteria down to
the $4\sigma$ limits of $y=26.1$ and $y=25.9$
in SDF and GOODS-N, respectively.
Note that the color criterion of $z'-y>1.5$ can be securely 
applied to our photometric catalogs 
down to the given $y$-band detection limits, 
because the $2\sigma$ upper limits of $z'$ images
reach $28.2$ (SDF) and
$27.4$ (GOODS-N).
Our candidate selection is made
with the photometric catalogs
of Suprime-Cam images alone
both in SDF and GOODS-N.
Although deep HST images are available
in the central $\sim 160$ arcmin$^2$ field of 
GOODS-N, we do not use these HST images
at this stage to avoid making a heterogeneous sample
given by the different detection limits
on the sky of GOODS-N. Instead, we take advantage of 
the deep HST images for confirming 
non-detections of blue continuum
for the candidates falling in
the area with the HST images (see below). 

\begin{figure}
\epsscale{2.3}
\plottwo{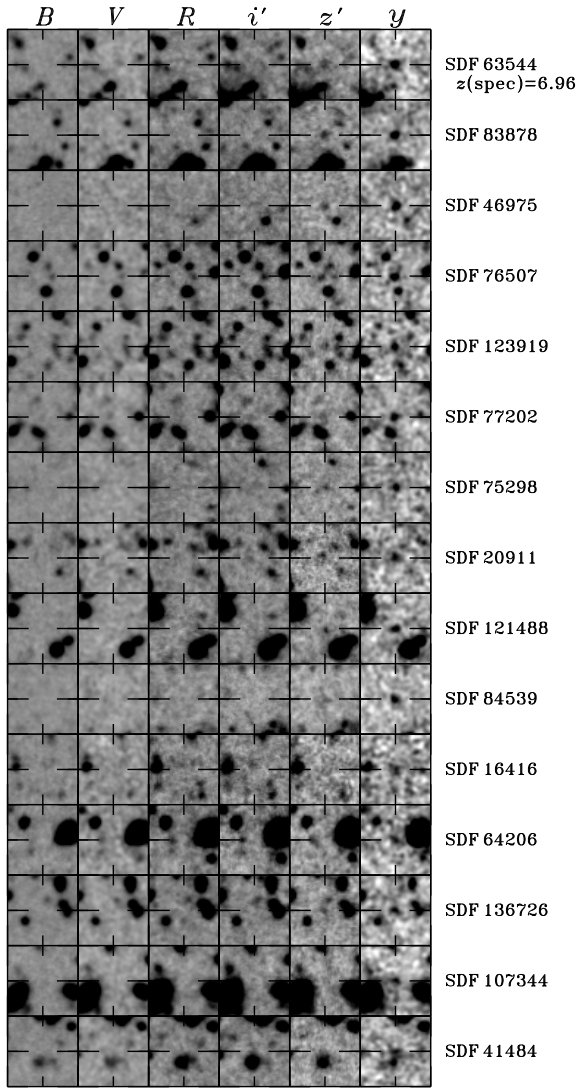}{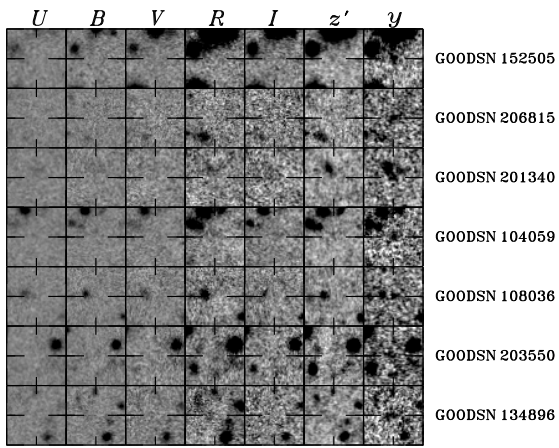}
\caption{
Snapshots of our $z$-dropout candidates 
identified in SDF (left) and GOODS-N (right). 
Each object has images of ($U$), 
$B$, $V$, $R$, $i'$, $z'$,
and $y$ bands with an ID number on the right.
The size of images is $10''\times 10''$.
North is up and east is to the left.
\label{fig:image_zdrop}}
\end{figure}

\begin{figure}
\epsscale{1.4}
\plotone{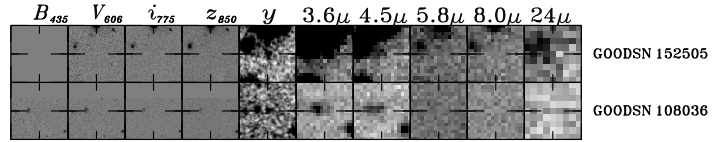}
\caption{
Snapshots of two $z$-dropout candidates falling 
in the area with the HST and Spitzer images in GOODS-N.
From left to right, we display HST/ACS $B_{435}$, 
$V_{606}$, $i_{775}$, $z_{850}$, Suprime-Cam $y$,
Spitzer/IRAC $3.6\mu$m, $4.5\mu$m, $5.8\mu$m, $8.0\mu$m, 
and Spitzer/MIPS $24\mu$m images. The Suprime-Cam $y$ image
is shown for comparison. GOODSN-152505 is not covered 
with the $B_{435}$ image.
The size of all images is $10''\times 10''$, but
that of $24\mu$m band is $11''\times 11''$.
North is up and east is to the left.
\label{fig:image_zdrop_goodsn_HSTSpitzer}}
\end{figure}

After we reject spurious sources near the spikes
of bright sources, the residuals of sky subtraction etc.
by visual inspection, we obtain $z$-dropout galaxy samples
consisting of 15 and 7 candidates in SDF and GOODS-N, 
respectively. All of these candidates have
magnitudes fainter than $y=25.4$ and brighter than
the detection limits in each field.
Snapshots of these 22 candidates are displayed in
Figure \ref{fig:image_zdrop}
and properties of our candidates 
are summarized in Table \ref{tab:sample}.
Some of our candidates are detected in $z'$- as well as $y$-bands.
Because the bandpass of $z'$ band includes both red and blue sides
of GP trough (1216\AA), the detections of faint $z'$ counterparts
are reasonable. The $z'$-band detections rather confirm
that the candidates are neither spurious sources nor transients 
appearing in the $y$ images.
Although we do not apply a criterion of non-detection 
in $i'$/$I$ band where a UV continuum 
between Lyman break (912\AA) and GP trough (1216\AA)
falls,
none of our candidates have an $i'$/$I$-band counterpart 
with a flux beyond our detection limits.

We check the spectroscopic catalogs of SDF and GOODS-N,
which are obtained by
\citet{kashikawa2003},
\citet{shimasaku2003},
\citet{ouchi2004},
\citet{shimasaku2006},
\citet{kashikawa2006},
\citet{yoshida2006}, \citet{iye2006}, \citet{nagao2007},
\citet{hayashi2009}, and \citet{ly2009} for SDF, and 
\citet{wirth2004}, \citet{reddy2006}, \citet{barger2008},
\citet{cohen2000,cohen2001}, \citet{steidel1996,steidel1999,steidel2003},
\citet{phillips1997}, \citet{lowenthal1997},
and \citet{dawson2001} for GOODS-N.
We find that one of our $z$-dropout candidates,
SDF-63544, has a spectroscopic redshift of $z=6.96$, which was 
originally identified by \citet{iye2006} in their 
Ly$\alpha$ emitter (LAE) study. SDF-63544 is the first dropout galaxy 
at $z\simeq 7$ with a spectroscopic redshift.
This object has a $y$-band magnitude
of $y=25.42$, which is the brightest candidate 
in our $z$-dropout galaxy samples.
It should be noted that our photometric sample 
surely includes a real $z\simeq 7$ galaxy.
On the other hand, we also find that none of our candidates
are low-$z$ objects with a spectroscopic redshift.
This confirms that our photometric criteria
do not select obvious foreground objects.

Two out of seven candidates in the GOODS-N field, 
GOODSN-152505 and GOODSN-108036, fall in
the region with the deep GOODS-N 
HST/ACS and Spitzer images.
We display snapshot images 
in Figure \ref{fig:image_zdrop_goodsn_HSTSpitzer}.
We have found that neither candidate
is detected in HST/ACS $B_{435}$, $V_{606}$,
and $i_{775}$ bands
\footnote{
GOODSN-152505 is located near the edge of
HST GOODS-N field, and not covered by the $B_{435}$ image.
}. 
The Spitzer images of GOODSN-152505 and GOODSN-108036 
are confused by the nearby bright objects 
due to large PSF sizes in IRAC and MIPS data.
However, there are some signatures of possible counterparts
for GOODSN-152505 in the $4.5\mu$m band
and for GOODSN-108036 in the $3.6\mu$m and $4.5\mu$m
bands, which would be different from 
the effects of source confusion.
Although photometry of these possible Spitzer counterparts are
more or less contaminated by fluxes of the nearby objects,
we simply calculate total magnitudes from $3''$-diameter 
aperture magnitudes and aperture corrections for IRAC and
MIPS fluxes given in \citet{yan2005} and the Spitzer web page
\footnote{
http://ssc.spitzer.caltech.edu/mips/apercorr/
}, 
respectively.
Since these sources are confused by the nearby objects,
we find that total magnitudes ($m_{\rm 3.6}$,$m_{\rm 4.5}$)
of GOODSN-152505 and GOODSN-108036 
are fainter than ($23.4$, $24.2$) and ($24.8$,$24.4$), respectively.
Thus, the colors of $y-m_{3.6}$ and $y-m_{4.5}$
are $\lesssim 1-2$ and $\lesssim 1$.
Because \citet{eyles2007} report that
their $z\sim 6$ galaxies with the IRAC detections
have colors of $1-2$ magnitudes in 
$z'-m_{3.6}$ and $z'-m_{4.5}$ 
(as well as $J-m_{3.6}$ $J-m_{4.5}$),
the colors of our two candidates 
are comparable to those of $z\sim 6$ galaxies.
This confirms that these two candidates are
not extremely red galaxies 
at $z\sim 2-3$, such reported by \citet{yan2004},
whose optical to IRAC colors ($z-m_{3.6}$) exceed 3.3.
Although the possible IRAC counterparts of our $z$-dropout
galaxy candidates would indicate that there exist
post-starburst galaxies even at $z\sim 7$,
there remains the possibility that these two candidates
are foreground interlopers. 
We will discuss stellar population of these two candidates
via detailed spectral energy distribution model fitting
after we confirm the redshifts of these candidates 
by spectroscopy.

\begin{figure}
\epsscale{1.11}
\plotone{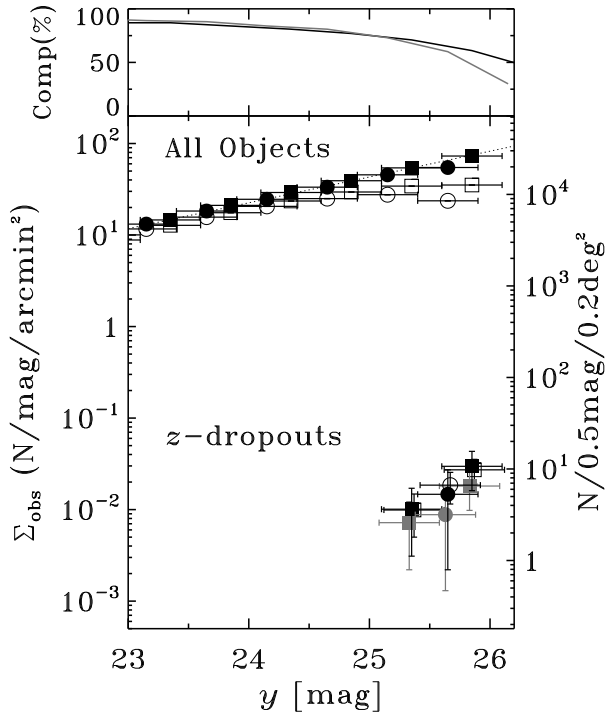}
\caption{
Top panel: Detection completeness of 
our $y$-band images in percentage. Black and gray 
solid lines represent the completeness
for $1''.8$-diameter aperture photometry
in SDF and GOODS-N, respectively.
Bottom panel: Surface densities of our objects
detected in our $y$ data. 
Lower and upper sequences of points 
show surface densities of our $z$-dropout galaxy candidates
and all objects detected in the $y$ images, respectively.
The surface densities are shown with the squares (SDF)
and circle (GOODS-N). In the upper sequence, 
filled black symbols are those with the completeness correction, 
while open symbols are not applied for the correction. 
In the lower sequence, filled gray symbols 
are surface densities that are subtracted with
the numbers of contamination, and filled black symbols
denote the best-estimate of our $z$-dropout galaxy surface densities
with both the contamination and completeness corrections (see \S \ref{sec:luminosity_function}).
We present total magnitudes from SExtractor's MAG\_AUTO
for the upper sequence to obtain the unbiased magnitude
surface density relation. Although the photometry of 
$z$-dropout galaxies is carried out with a $1''.8$-diameter 
aperture to maximize the signal-to-noise ratio of faint
sources, the aperture corrections are applied for 
$z$-dropout galaxies in the luminosity function plot of 
Figure \ref{fig:lumifun_UV_diff_zdrop}.
For the presentation purpose, we slightly shift 
the open and gray points along the abscissa.
The exact magnitudes are the same as magnitudes of 
black filled points. The vertical axis on the right side
indicates the number counts of objects, 
i.e. $N/(0.5 {\rm mag})/(0.2{\rm deg}^2)$,
which approximately correspond to numbers of $z$-dropout galaxies
identified in each target field.
Dotted line represents the best-fit power law
to the completeness-corrected surface densities of all objects.
\label{fig:number_density_zr_pluscomp}}
\end{figure}

There are no obvious counterparts of these two candidates 
at longer wavelengths; IRAC $5.8\mu$m, $8.0\mu$m, 
and MIPS $24\mu$m bands.
The MIPS snapshot of GOODSN-152505 shows 
a source at the left side, but 
this MIPS source is a counterpart of 
a bright source located near the left corner.
Because the detection limits of the three bands
are too shallow ($21-23$ mag; Table \ref{tab:limitingmag}) 
to identify galaxies at large distances,
the non detections, again, confirm that 
these two candidates are neither
extremely red galaxies nor AGN at low redshifts.

\section{Luminosity Function}
\label{sec:luminosity_function}

\subsection{Surface Number Densities and Detection Completeness}
\label{sec:number_density}

We obtain the number counts
of all $y$-band detected objects, $N_{\rm all}(m)$,
and our $z$-dropout galaxy candidates, $N_{\rm cand}(m)$,
from our photometric catalogs. 
We calculate the surface number densities, $\Sigma_{\rm obs}$,
by dividing $N_{\rm all}(m)$ and $N_{\rm cand}(m)$ by our respective survey areas. 
The results are presented in the bottom panel of Figure \ref{fig:number_density_zr_pluscomp}. 
Since the surface number densities of faint objects are affected by detection incompleteness, 
we estimate detection completeness as a function of $y$ magnitude
by Monte Carlo simulations.
We distribute 7240 artificial objects with a point spread function
on our $y$-band images after adding photon Poisson noise,
and detect them in the same manner as
for the detection for our photometric catalogs with SExtractor.
We repeat this process 20 times, and compute the ratio of
recovered objects to the input objects. 
The top panel of Figure \ref{fig:number_density_zr_pluscomp} 
shows the detection completeness of our $y$-band images.
We find that the detection completeness is typically
$\gtrsim 70$\% for relatively luminous sources with $y\lesssim 25.5$.
The detection completeness is $>50$\% even in the faintest magnitude 
bins centered at $y=25.85$ (SDF) and $y=25.65$ (GOODS-N).
We correct the surface number densities for the detection completeness,and present them in 
the bottom panel of Figure \ref{fig:number_density_zr_pluscomp}.

\subsection{Contamination}
\label{sec:contamination}
There are four sources of contamination in our $z\sim 7$ galaxy samples:
1) spurious $y$-band sources made of noise fluctuations,
2) transients, such as faint variable stars+AGN and supernovae detected 
in our $y$-band images,
3) foreground red objects entering our samples due to photometric errors,
4) L/T dwarf stars satisfying our color selection criteria.
We define the numbers of contamination for 1), 2), 3), and 4) in our samples 
as
$N_{\rm cont}^1(m)$, $N_{\rm cont}^2(m)$, $N_{\rm cont}^3(m)$, and
$N_{\rm cont}^4(m)$, respectively.
Below, we check the effects of contamination, and estimate
their impacts on our $z$-dropout galaxy samples.

1) Spurious sources:

Because we push our $y$-band detection limits,
our $z\sim 7$ galaxy samples may include spurious sources made of noise peaks. 
Some of these $y$-band spurious sources could pass our selection criteria
of eq. (\ref{eq:zdropselection}). 
To estimate how much spurious sources are included in
our $z$-dropout galaxy samples,
we carry out source detection and color selection same as 
those in \S \ref{sec:samples},
but with images whose ADU counts are multiplied by $-1$. We run SExtractor
with these negative-count images, and make negative-$y$ band detection catalogs.
We apply the color criteria of eq. (\ref{eq:zdropselection}), and 
reject sources apparently made by the residuals of sky subtraction
in the same manner as for the real $z$-dropout galaxy selection.
We find 0 and 1 spurious $z\sim 7$ galaxy candidate
in the negative SDF and GOODS-N images, respectively.
We conclude that our samples include $N_{\rm cont}^1(m)= 0$ (SDF) 
and $1.0\pm 1.0$ (GOODS-N) spurious source statistically,
where the error is estimated by Poisson statistics.

2) Transients:

Although our $y$ images are 
the stacked data that were acquired
in a 6-year (3-year) long period
from 2003 (2006) to 2009 for SDF (and GOODS-N;
see Table \ref{tab:obs}), we investigate
the possibility of transients for our $z$-dropout
candidates.
We stack $y$ data taken before
and after 2008, and obtain $y$-band images
for the two epochs. 
The $y$-band detection limits of the former ($y_{\rm lim}^{\rm epoch1}$) 
and the latter ($y_{\rm lim}^{\rm epoch2}$) images are 
$(y_{\rm lim}^{\rm epoch1},y_{\rm lim}^{\rm epoch2})=$
$(25.9,26.2)$ in SDF and
$(26.0,25.6)$ in GOODS-N.
We have carried out photometry at the positions of 
our $z$-dropout galaxy candidates, and found no candidates
detected at the $3\sigma$ levels
that show a significant magnitude change
between these two epochs by $\gtrsim 0.5$ magnitude 
(corresponding to $1-2$ sigma levels).
Thus, we conclude $N_{\rm cont}^2(m)=0$ in both SDF and GOODS-N.
Checking this result, we calculate an expected number of transients
based on the deep and wide-field transient study results
of \citet{morokuma2008}. \citet{morokuma2008} present
that the number density of transients (with the timescale greater than 200 days)
in the magnitude range of $i'\simeq 25.4-26.0$ is $\simeq 0.4$/deg$^2$ 
Assuming the difference of magnitudes between $y$ and $i'$ is negligible
for transients which are mostly nearby objects, the expected number of transients
in our SDF and GOODS-N samples
is only $\sim 0.1$ with the detection completeness correction.
This estimate is consistent with our conclusion of no transients 
in our $z$-dropout galaxy samples.

3) Foreground red objects entering our samples due to photometric errors:

It is possible that some foreground objects,
such as red galaxies at intermediate redshifts, 
enter our color criteria by photometric errors,
although their intrinsic colors cannot satisfy
the criterion of $z-y>1.5$. 
We make an input mock catalog mimicking
foreground objects, and carry out Monte Carlo 
simulations with the mock catalog 
to estimate the numbers of foreground interlopers.
The mock catalog has the same number-density distribution
as that of all $y$-detected objects corrected for
the detection completeness (Filled circles at
the upper sequence in the bottom panel of Figure \ref{fig:number_density_zr_pluscomp}). 
In the mock catalog, colors of the bright ($y<24.5$) objects 
are the same as those of the observed $y$-detected objects.
Because faint objects have moderately large photometric
errors, we do not use colors of objects with a magnitude
fainter than $y=24.5$. Instead, we assign color distribution of 
observed $y$-detected objects with $y=23.5-24.5$ to the objects 
with $y>24.5$ in the mock catalog.
Then, we perform Monte-Carlo simulations
in the same manner as \S \ref{sec:number_density};
distributing artificial objects from the mock catalog
on our real images and detecting them with SExtractor.
We find that expected numbers of these interlopers are
$N_{\rm cont}^3(m)=0.85$ in SDF down to $y=26.1$,
and 
$0.80$ in GOODS-N down to $y=25.9$.

4) L/T dwarf stars:

Figure \ref{fig:redshift_color_paper} indicates that
none of foreground objects except Galactic late-type
stars can satisfy our color criterion of $z-y>1.5$
without photometric errors. Late-type
stars are potentially the largest source of contamination
in our $z$-dropout galaxy samples.
Estimating the numbers of late-type stars
which contaminate our $z$-dropout samples, 
we carry out Monte-Carlo simulations same as 3),
but with an input mock catalog of late-type stars.
We use the number density of L/T dwarfs as a function
of Galactic latitude presented in \citet{ryan2005} 
who derive the number densities in 15 deep HST/ACS fields
down to $z=26.0$ at various Galactic latitude and
fit them to the model of surface density averaged 
over Galactic longitude. The colors of L/T dwarfs 
are calculated with the spectral templates of \citet{knapp2004} 
that cover L1-L9 and T0-T9 stars.
The results of these Monte-Carlo simulations
show that the numbers of late-type star contaminants
are $N_{\rm cont}^4(m)=5.35$ down to $y=26.1$ in SDF,
and $2.80$ down to $y=25.9$ in GOODS-N.
The uncertainties of these simulation results
are $0.4-0.5$ in number.

\begin{figure}
\epsscale{1.15}
\plotone{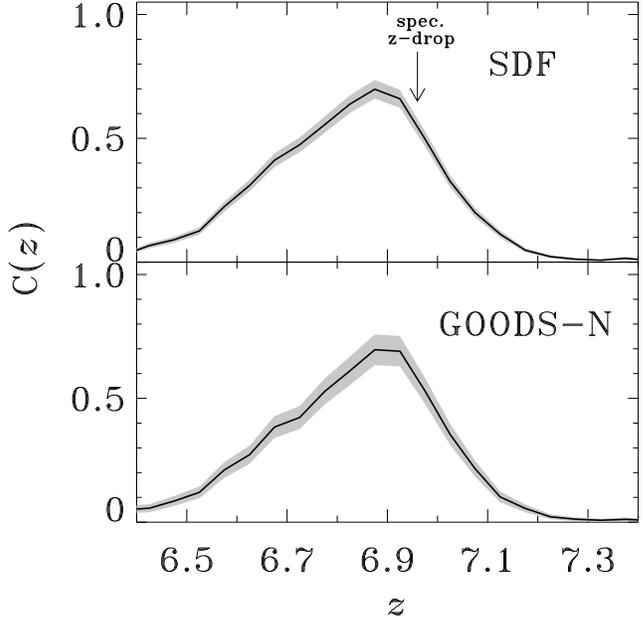}
\caption{
Redshift distribution of our $z$-dropout galaxies
in SDF (top) and GOODS-N (bottom).
Solid lines plot
the redshift distributions
averaged over magnitudes
weighted with the number density distributions of 
our $z$-dropout galaxies.
Gray shades represent errors of
the redshift distribution estimates
obtained by our Monte-Carlo simulations.
Arrow indicates the redshift of our $z$-dropout galaxy 
with the spectroscopic confirmation (SDF-63544; $z_{\rm spec}=6.96$).
\label{fig:redshift_comp}}
\end{figure}

\subsection{Redshift Distribution}
\label{sec:redshiftdistribution}

We have estimated redshift distribution of our $z$-dropout galaxies, $C(m,z)$,
by Monte-Carlo simulations with an input mock catalog of high-$z$ galaxies.
The mock catalog consists of high-$z$ galaxies whose properties
are given with the probability distributions of 
i) number count, ii) continuum color,
and iii) Ly$\alpha$ emissivity.
First, for the probability distribution of i),
we use the surface number densities of our $z$-dropout galaxies 
corrected for contamination and completeness 
(black filled points in Figure \ref{fig:number_density_zr_pluscomp}).
Second, we assume that high-$z$ galaxies have the
average UV continuum slope of $\beta=-2$ which 
is found in $z\sim 6$ dropout galaxies 
\citep{stanway2005,yan2005,bouwens2006,bouwens2007,bouwens2009c},
and that the probability distribution of $\beta$
is a Gaussian function with a standard deviation of $\sigma_{\beta}=0.5$.
We generate spectra with the stellar population synthesis model of 
\citet{bruzual2003}, and obtain galaxy spectra with 
$\beta=(-3.0)-(-1.0)$. The model parameters are the
same as those of the average values estimated for
Lyman break galaxies (LBGs) at $z=3$ \citep{papovich2001}, but
with a young age of $4$ Myr and Calzetti dust extinction
ranging from $E(B-V)=0.008$ to $0.400$, so that we can 
reproduce a set of spectra covering from the very blue ($\beta=-3.0$) 
to the moderately red ($\beta=-1.0$) continua.
We apply a GP optical depth calculated from 
\citet{madau1995} for the IGM absorption, and 
use this probability distribution for ii).
Third, for the fraction of Ly$\alpha$ emitting 
galaxies, we assume that 30\% of $z=7$ dropout galaxies have
a Ly$\alpha$ emission line with a rest-frame
equivalent width ($EW_0$) of $>20$\AA,
which is the same as the estimates for $z=6$ dropout galaxies
down to $\sim L^*$ 
\citep{stanway2004a,stanway2004b,vanzella2006a,dow-hygelund2007,stanway2007}. 
We add a Ly$\alpha$ luminosity
to these 30\% of dropout galaxies with the $EW_0$ probability distribution of 
LAEs at $z=5.7$ derived by \citet{ouchi2008}.
Since \citet{shapley2003} have found that 
about 50\% of their $z=3$ LBGs have a Ly$\alpha$ $EW_0$ of $\le 0$\AA,
we assume that a half of our $z=7$ dropout galaxies
have no Ly$\alpha$ emission line. For the rest of 20\% of
high-$z$ galaxies, we add a very weak $EW_0=0-20$
Ly$\alpha$ line to their spectra. 

Finally, we produce an input mock catalog of high-$z$ galaxies
that are distributed in the radshift space homogeneously,
and apply statistical weights following the probability distributions of 
i), ii), and iii) shown above. 
This mock catalog is used to carry out
Monte-Carlo simulations in the same manner as 
those in \S \ref{sec:contamination}.
Note that luminosity function around 
the detection limits are usually overestimated 
because Poisson noise produces more sources with an 
up scattered flux than those with a down scattered 
flux at a given flux bin for objects with a typical 
steep source number count \citep{eddington1940}.
To correct this bias, we extrapolate 
the relation between surface number density and 
magnitude down to $y=27$, and produce the mock catalog
including faint sources below our detection limit.
We obtain $y$-band detection catalogs
by the Monte-Carlo simulations, and 
select artificial $z$-dropout galaxies
with the color criteria (eq. \ref{eq:zdropselection})
to draw mock samples of $z$-dropout galaxies. 
We calculate the ratio of the selected objects
to the input objects as a function of redshift, which corresponds
to the redshift distribution of our $z$-dropout galaxies.
Figure \ref{fig:redshift_comp} plots 
the redshift distribution of our $z$-dropout galaxies, $C(z)$,
averaged over magnitudes with 
the probability distributions of i).
For both data of SDF and GOODS-N, the peak redshift of $C(z)$ is 
$z=6.9$, and 90 percent of the $z$-dropout galaxies fall in $z=6.5-7.1$.
Thus, the redshift window of our $z$-dropout selection
is $z=6.9^{+0.2}_{-0.4}$.

\subsection{UV Luminosity Function}
\label{sec:uv_luminosity_function}

We derive the UV luminosity function of $z$-dropout galaxies 
based on the numbers of our candidate galaxies (\S \ref{sec:number_density})
and the contamination objects (\S \ref{sec:contamination}),
and the completeness as a function of redshift, i.e. redshift
distribution (\S \ref{sec:redshiftdistribution}).
We calculate the number density, $n(m)$, of $z$-dropout galaxies
at each field in a given magnitude bin by
\begin{equation}
n(m)=\frac{\left[N_{\rm cand}(m)-\sum_{i=1}^{n} N_{\rm cont}^i(m)\right]}{\int_{0}^{\infty} \frac{dV}{dz} C(m,z) dz},
\label{eq:numberdensity}
\end{equation}
where $n$ represents the four kinds of contaminants satisfying 
our selection criteria in a given magnitude bin ($n=4$; see \S \ref{sec:contamination}),
and $\frac{dV}{dz}$ is the differential cosmic volume with an area of SDF or GOODS-N.
Because we apply a $1''.8$-diameter aperture
for $z$-dropout galaxy photometry to maximize 
the signal-to-noise ratio, we need to brighten 
the $y$ magnitudes with an aperture correction 
to estimate total fluxes. 
On the other hand, our $y$-band magnitudes can be
contaminated with Ly$\alpha$ emission lines.
We should subtract the contributions of Ly$\alpha$ fluxes
to obtain UV-continuum magnitudes, and dim 
the $y$-band brightness accordingly. Moreover, we should apply
k-correction to get UV-continuum luminosities 
at the rest-frame $\sim 1500$\AA\ from 
our $1200-1300$\AA\ luminosities including
small portions of Ly$\alpha$ forest below GP trough,
which enter the bandpass of our $y$ filter.
We estimate a correction factor to derive
total UV-continuum magnitudes 
at $\sim 1500$\AA\ from our $y$-band magnitudes
based on the results of our Monte-Carlo simulations in
\S \ref{sec:redshiftdistribution}. 
We calculate the mean input 1500\AA-continuum
and the output $y$-band magnitudes averaged 
with the statistical weights from 
the probability distribution functions 
(\S \ref{sec:redshiftdistribution}), 
and obtain correction magnitudes of $+0.01$ and $+0.05$ 
for SDF and GOODS-N samples, respectively. 
We apply these small corrections to 
our UV-continuum magnitude estimates

\begin{figure}
\epsscale{1.2}
\plotone{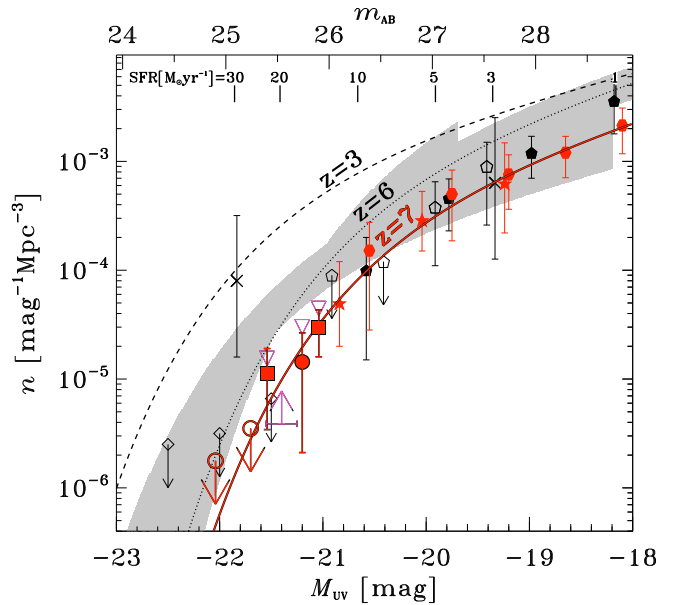}
\caption{
UV luminosity function (LF) of $z$-dropout galaxies,
together with those at lower redshifts.
Two red filled squares and one red filled circle 
present UV LF of our $z$-dropout galaxies 
in SDF and GOODS-N, respectively.
Two red arrows with open circles indicate 
upper limits of $z$-dropout galaxies estimated 
from our GOODS-N data (right) and 
the combination of our SDF and GOODS-N data (left).
Magenta arrow displays the lower limit estimated
from our spectroscopically-identified $z$-dropout galaxy.
Magenta inverse-triangles represent the maximal LF,
i.e., the secure upper limits of our LF estimates 
that include no correction for contamination.
Recent measurements and upper limits
including those from HST/WFC3 studies
are also plotted with 
red star marks \citep{bouwens2008}, 
hexagons \citep{mclure2009b}, 
black filled pentagons \citep{oesch2009b}, 
open pentagons \citep{oesch2009a}, 
open diamonds \citep{mannucci2007}, 
and crosses \citep{richard2006}.
Although previous studies define
$M_{\rm UV}$ at the rest-frame $\simeq 1300-2000$\AA,
a k-correction between $\simeq 1400$\AA\ and $2000$\AA\ is
as small as $\sim 0.07$ \citep{oesch2009a}.
Red solid line plots our best-fit Schechter function
of $z$-dropout galaxies at $z=7$. 
Dotted line is UV LF at $z\sim 6$ \citep{bouwens2008}.
Gray shade denotes UV LF from the other various studies 
(see \citealt{ouchi2008} and the references therein).
Dashed line is UV LF at $z\sim 3$ \citep{steidel1999}.
Ticks on the upper horizontal axis show
observed broad-band magnitudes of objects at $z=7$. 
At around the top of this plot,
we also tick the corresponding star-formation rates estimated
from eq. (\ref{eq:uv_sfr}) with no dust extinction correction.
\label{fig:lumifun_UV_diff_zdrop}}
\end{figure}

Figure \ref{fig:lumifun_UV_diff_zdrop} presents
UV LF of $z$-dropout galaxies from our samples with the red filled squares
(SDF) and circle (GOODS) as well as our upper limits with the red open circles.
We also plot $z$-dropout galaxy UV LFs derived 
by the deep HST NICMOS+WFC3
\citep{bouwens2008,oesch2009a,oesch2009b,mclure2009b} and 
the ground-based \citep{richard2006,mannucci2007} studies, together with 
UV LFs at low redshifts. 
\footnote{
We cannot include the recent results of
\citet{bunker2009}, \citet{castellano2009}, and \citet{hickey2009},
because their UV LF measurements are not apparently presented.
}
Because we have a spectroscopically
identified galaxy at $z=6.96$, we can also place a lower limit
on the UV LF. We estimate a 1500\AA-continuum magnitude
of this galaxy to be $M_{\rm UV}=-21.40\pm 0.31$
from the y-band photometry ($y=25.42$), Ly$\alpha$ flux 
($2\times 10^{-17}$ erg s$^{-1}$ cm$^{-2}$; \citealt{iye2006,ota2008}), 
and the mean model spectrum at $z=6.96$ with $\beta=-2.0$ 
(\S \ref{sec:redshiftdistribution}).
We show this lower limit with the magenta arrow 
in Figure \ref{fig:lumifun_UV_diff_zdrop}.

Note that our study provides, for the first time, 
measurements of $z=7$ UV LF at 
the bright magnitude of $-22<M_{\rm UV}<-21$
as well as the lower limit based on
the spectroscopically-identified $z$-dropout galaxy.
We find that our UV LF at $z=7$ 
falls significantly below that at $z=6$.
Because our $z$-dropout galaxy samples are largely corrected for 
contamination estimated by the simulations
(\S \ref{sec:contamination}),
these small number densities of 
our $z$-dropout galaxies could be due to over estimates of 
the contaminants.
We derive UV LF of our $z$-dropout galaxies in the same manner as above,
but with no contamination subtraction, and refer to 
these estimates as the maximal LF which provides
conservative upper limits.
We show the maximal LF with the magenta inverse-triangles 
in Figure \ref{fig:lumifun_UV_diff_zdrop}.
These magenta inverse-triangles also fall below the
$z=6$ UV LF measurements.
Note that the $z=6$ UV LF measurements would also have
the similar problems in contamination estimates,
but our magenta inverse-triangles come below
the gray shade area in Figure \ref{fig:lumifun_UV_diff_zdrop}
that represents a variance of $z=6$ LFs 
derived by different studies 
with various contamination estimates.
Thus, we conclude 
with our LF measurements and maximal LF estimates 
that $z=7$ UV LF definitely decreases 
from $z=6$ at the bright end. 
Our conclusion is
consistent with the claim of \citet{mannucci2007,castellano2009}.
Moreover, this decreasing tendency is similar to
that found at the faint magnitudes
by the HST studies ($M_{\rm UV}>-21$; 
\citealt{bouwens2008,oesch2009a,oesch2009b,mclure2009b,bunker2009}).
The combination of our results and the HST studies
suggests that UV LF decreases at both bright and 
faint magnitudes from $z=6$ to $7$.
We fit a Schechter function to the LF measurements
from our and the HST studies, and obtain the best-fit parameters of 
$\phi^*=0.69_{-0.55}^{+2.62} \times 10^{-3}$Mpc$^{-3}$,
$M_{\rm UV}^* = -20.10\pm 0.76$ mag, and
$\alpha = -1.72\pm 0.65$ 
that maximize the likelihood,
${\cal L}=
\Pi_{i}\ p[N_{\rm obs}(m_i),N_{\rm exp}(m_i; \phi^*, M_{\rm UV}^*, \alpha)]$,
where $p[x, \mu]$ is the Gaussian distribution with a mean $\mu$
evaluated at $x$, and $N_{\rm obs}$ and $N_{\rm exp}$ are, respectively, 
the numbers of galaxies within a magnitude bin of $m_i$ 
from observations and expectations
for a given set of Schechter parameters.
\footnote{
To avoid using the dependent measurements from the same HST data,
we exclude the results of \citet{oesch2009a,oesch2009b}
in the fitting. The inclusion of \citet{oesch2009a,oesch2009b}
data does not change our conclusions but with 
artificially small errors.
}
We summarize the best-fit Schechter parameters 
in Table \ref{tab:muv_lumifun_schechter},
and plot the best-fit function
in Figure \ref{fig:lumifun_UV_diff_zdrop} with the red line.
Although the constraints on
$\alpha$ are weak, a steep slope of $\alpha$ is suggestive,
which is similar to that at $z\lesssim 6$ \citep{bouwens2008,mclure2009a}.

Since our survey results are based on two fields, SDF and GOODS-N,
which are well-separated on the sky, we can estimate
the effect of the cosmic variance
on the number densities of $z$-dropout galaxies. 
First, the numbers (+Poisson errors)
of $z$-dropout galaxies
in samples of (SDF, GOODS-N) are 
($2\pm 1.4$, $1\pm 1.0$) at $y<25.5$,
($4\pm 2.0$, $3\pm 1.7$) at $y<25.7$, and
($8\pm 2.8$, $7\pm 2.6$) at $y<25.9$
\footnote{
Seven ($=15-8$) candidates of SDF fall in a narrow magnitude
window of $y=25.9-26.1$. Because 
the test in \S \ref{sec:contamination} 
indicates that our samples include $0-1$ spurious sources.
Thus, the moderately large number of faint candidates in SDF
is not due to spurious sources 
but a steep slope of number count for $z$-dropout galaxies
at this magnitude regime.
}.
Since the survey areas of SDF and GOODS-N
are comparable ($\simeq 0.2$ deg$^2$; 
\S \ref{sec:observations_and_data_reduction}), 
the number counts in these two fields
are comparable within the errors.
Second, we compare the LF measurements
of SDF and GOODS-N in Figure \ref{fig:lumifun_UV_diff_zdrop},
and find that the cosmic variance is smaller than a factor of $\sim 2$ 
in number density for the survey areas of $0.2$ deg$^2$.
We estimate the expected cosmic variance with the analytic CDM model of
\citet{sheth1999} from the number density of our $z$-dropout galaxies
down to our magnitude limit ($0.9_{-0.6}^{+2.2} \times 10^{-5}$ Mpc$^{-3}$ 
at $M_{\rm UV}<21$), assuming one-to-one correspondence between
galaxies and dark halos. The corresponding clustering bias
is $b=7.6-9.6$, and the cosmic variance is 22-28\%
for an area of 0.2 deg$^2$ at $z=6.5-7.1$.
We check our calculations with the Cosmic Variance Calculator
\footnote{
http://solo.colorado.edu/~trenti/CosmicVariance.html
}
\citep{trenti2008}, and find that this calculator
returns a very comparable number of $\simeq 30$\% 
for the cosmic variance after subtraction of 
Poisson error term.
We obtain even a smaller cosmic variance of $\simeq 9-12$\%
with the CDM model of \citet{sheth1999}, if we do not assume
the one-to-one correspondence, but adopt clustering bias of
galaxies measured at a slightly lower redshift of $z\sim 6$ 
($b\simeq 3-4$; \citealt{ouchi2005,overzier2006}). 
Thus, the prediction of CDM model is consistent with
our observational constraints of the cosmic variance, $\lesssim 2$,
in the two areas of 0.2 deg$^2$.

\begin{figure}
\epsscale{1.2}
\plotone{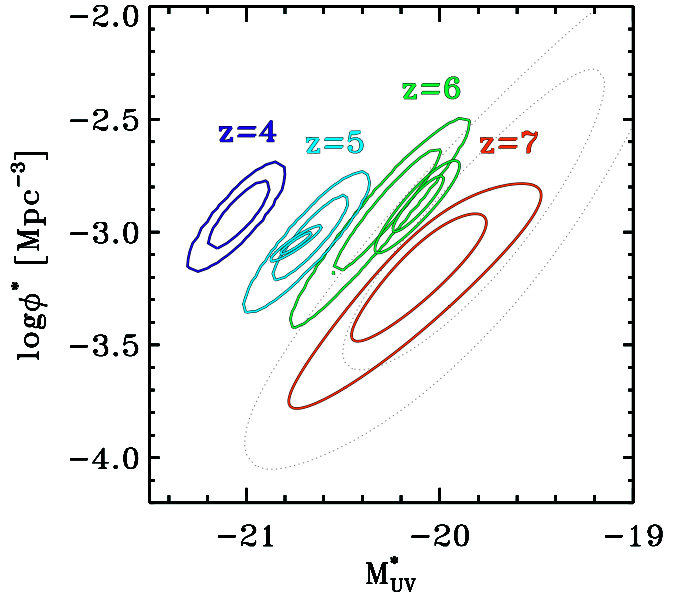}
\caption{
Error ellipses of Schechter parameters,
$M^*$ and $\phi^*$, 
at the $1$ and $2\sigma$ confidence levels.
Red lines represent our results of
$z$-dropout galaxies at $z=7$.
Blue, cyan, and green contours
denote error ellipses for galaxies at 
$z=4$, $5$, and $6$ obtained by
\citet{bouwens2008} (large contours)
and \citet{mclure2009a} (small contours 
only for $z=5$ and $6$). The dotted contours
are error ellipses of $z=7$ galaxies estimated
by \citet{bouwens2008}. All of these Schechter
fits are based on $\alpha \simeq -1.7$.
Our measurements of the red contours 
indicate that the Schechter parameters  
of $z=7$ LF differ from those of $z\le 6$ LFs
at the $>2\sigma$ (i.e. $>95$\%) level.
\label{fig:contour_schechter_z7drop}}
\end{figure}

Figure \ref{fig:contour_schechter_z7drop}
shows the error ellipses of our Schechter parameters,
$M^*$ and $\phi^*$, for $\alpha=-1.72$ at the $1$ and $2\sigma$ confidence levels. 
We also present those of LFs at $z=6$ and $5$ \citep{bouwens2008,mclure2009a}
and $z=4$ \citep{bouwens2008}. Note that all of these Schechter fits 
are based on $\alpha \simeq -1.7$ ($\alpha=[-1.66]-[-1.74]$ for \citealt{bouwens2008}
and $\alpha=[-1.66]-[-1.71]$ for \citealt{mclure2009a}), and that 
Figure \ref{fig:contour_schechter_z7drop} exclusively compares two parameters of 
$M^*$ and $\phi^*$.
Our measurements (red contours in Figure \ref{fig:contour_schechter_z7drop}) 
indicate that the Schechter parameters  
of $z=7$ LF differ from those of $z\le 6$ LFs
significantly, and that LF decreases from $z=6$ to $7$
at more than the $2\sigma$ (i.e. $95$\%) level.
Moreover, our constraints of $z=7$ Schechter parameters 
are consistent with those of \citet{bouwens2008} (dotted lines
in Figure \ref{fig:contour_schechter_z7drop}), but are stronger
than those, which allow us to rule out
no evolution at the $>95$\% level. 
Although the errors of 
our measurements are too large to distinguish between
luminosity ($L^*$) and number ($\phi^*$) evolutions,
Figure \ref{fig:contour_schechter_z7drop} implies
that a decrease in $L^*$ would be the dominant factor
of the LF evolution from $z=5-6$ to $7$.

\subsection{UV Luminosity Density}
\label{sec:uv_luminosity_density}

We calculate UV-luminosity densities at $z=7$
from our UV LF with the best-fit Schechter parameters given 
in \S \ref{sec:uv_luminosity_function}.
First, we integrate the LF down to the magnitude of 
the faintest observed $z$-dropout galaxies of the HST/WFC3 studies,
which was used for our Schechter parameter fitting 
(i.e., down to $M_{\rm UV}=-18$ or $\simeq 0.1 L^*$),
and obtain the observed UV-luminosity density, $\rho_{\rm UV}^{\rm obs}$.
Because a total UV-luminosity density has to be larger than
$\rho_{\rm UV}^{\rm obs}$ by the amount of the contribution 
from galaxies fainter than the limiting magnitude, 
$\rho_{\rm UV}^{\rm obs}$ corresponds to the lower limit of 
the UV-luminosity density. We extrapolate our LF 
down to $L=0$ to estimate the contribution from such faint galaxies,
and obtain the UV-luminosity density, $\rho_{\rm UV}^{\rm upper}$.
Since there exist no galaxies with $L\sim 0$, this UV-luminosity density with
the LF extrapolation, $\rho_{\rm UV}^{\rm upper}$, corresponds to 
the upper limit of UV luminosity density.
We estimate 
$\rho_{\rm UV}^{\rm obs} = 4.0_{-2.6}^{+6.2} \times 10^{25}$ erg s$^{-1}$ Hz$^{-1}$Mpc$^{-3}$
and 
$\rho_{\rm UV}^{\rm upper} = 1.1_{-0.5}^{+1.1} \times 10^{26}$ erg s$^{-1}$ Hz$^{-1}$Mpc$^{-3}$.
Both $\rho_{\rm UV}^{\rm obs}$ and $\rho_{\rm UV}^{\rm upper}$ 
are summarized in Table \ref{tab:muv_lumifun_schechter}.
We discuss evolution of cosmic star-formation rate density 
(\S \ref{sec:cosmic_starformation_history}) and 
reionization of the Universe 
(\S \ref{sec:reionization}) using 
these UV-luminosity densities.
For the upper limit value, there are uncertainties in
the faint-end slope of the LF, $\alpha=-1.72$, 
which we have weakly constrained. 
If we assume the best-fit Schechter parameters 
but with a steep slope of $\alpha=-1.86$, 
$\rho_{\rm UV}^{\rm upper}$ would increase 
only by a factor of 2, which just corresponds to 
the $1\sigma$-upper error value ($2.2\times 10^{26}$erg s$^{-1}$ Hz$^{-1}$Mpc$^{-3}$) 
of our $\rho_{\rm UV}^{\rm upper}$.
On the other hand, a very steep slope of
$\alpha=-1.90$ could push up $\rho_{\rm UV}^{\rm upper}$ by 
a factor of 3, and an extreme value of $\alpha=-1.97$ 
may boost $\rho_{\rm UV}^{\rm upper}$ by a factor of 10.
However, \citet{bouwens2008} and \citet{mclure2009a} have 
reported $\alpha$ measurements similar to ours
but with a small uncertainty, $\alpha=-1.74 \pm 0.16$ and
$\alpha=-1.71 \pm 0.11$, respectively, 
for galaxies at a slightly lower redshift of
$z=6$. Since it is unlikely that the faint-end slope
evolves largely between $z=6$ to $7$, 
$\rho_{\rm UV}^{\rm upper}$ would not be 
well beyond its $1\sigma$-upper error value
in the reasonable range of $\alpha$.

\section{Discussion}
\label{sec:discussion}

\subsection{Cosmic Star-Formation History}
\label{sec:cosmic_starformation_history}

We calculate cosmic star-formation rate densities (SFRDs) from the
UV-luminosity densities, $\rho_{\rm UV}^{\rm obs}$ and 
$\rho_{\rm UV}^{\rm upper}$. 
We use the relation between UV luminosity 
and star-formation rate (SFR) given by \citet{madau1998}:
\begin{equation}
{\rm SFR}(M_\odot \ {\rm yr}^{-1}) = 
L_{UV}{\rm (erg\ s^{-1} Hz^{-1})}/(8\times 10^{27}),
\label{eq:uv_sfr}
\end{equation}
where $L_{\rm UV}$ is UV luminosity
measured at 1500\AA.
This relation assumes that galaxies have 
the Salpeter IMF with solar metallicity.
We obtain ${\rm SFRD}^{\rm obs}= 4.95_{-3.24}^{+7.75} \times 10^{-3}$ $M_\odot$ yr$^{-1}$ Mpc$^{-3}$ and 
${\rm SFRD}^{\rm upper}=  1.32_{-0.66}^{+1.35} \times 10^{-2}$ $M_\odot$ yr$^{-1}$ Mpc$^{-3}$ from 
$\rho_{\rm UV}^{\rm obs}$ and 
$\rho_{\rm UV}^{\rm upper}$, respectively.

We apply extinction correction to the SFRDs, assuming
the empirical relation between the UV slope, $\beta$, 
and extinction, $A_{1600}$, for starburst galaxies,
\begin{equation}
A_{1600}=4.43+1.99\beta
\label{eq:extinction}
\end{equation}
\citep{meurer1999}.
Following the observational results of $z\sim 6$ dropout galaxies
\citep{stanway2005,yan2005,bouwens2006,bouwens2007,bouwens2009c},
we use $\beta=-2$ for our $z$-dropout galaxies.
We estimate the extinction-corrected SFRDs to be
${\rm SFRD}^{\rm obs}_{\rm corr}= 7.49_{-4.90}^{+11.7} \times 10^{-3}$ $M_\odot$ yr$^{-1}$ Mpc$^{-3}$ and 
${\rm SFRD}^{\rm upper}_{\rm corr}= 2.00_{-1.00}^{+2.04} \times 10^{-2}$ $M_\odot$ yr$^{-1}$ Mpc$^{-3}$.
Since ${\rm SFRD}^{\rm upper}_{\rm corr}$ is the SFRD with dust correction and
extrapolation of the LF down to $L=0$, ${\rm SFRD}^{\rm upper}_{\rm corr}$ is
an upper limit of our SFRD measurements. On the other hand,
${\rm SFRD}^{\rm obs}$ is the SFRD estimated with neither dust extinction correction 
nor LF extrapolation. Thus, ${\rm SFRD}^{\rm obs}$ is regarded as a conservative
lower limit.

\begin{figure}
\epsscale{1.2}
\plotone{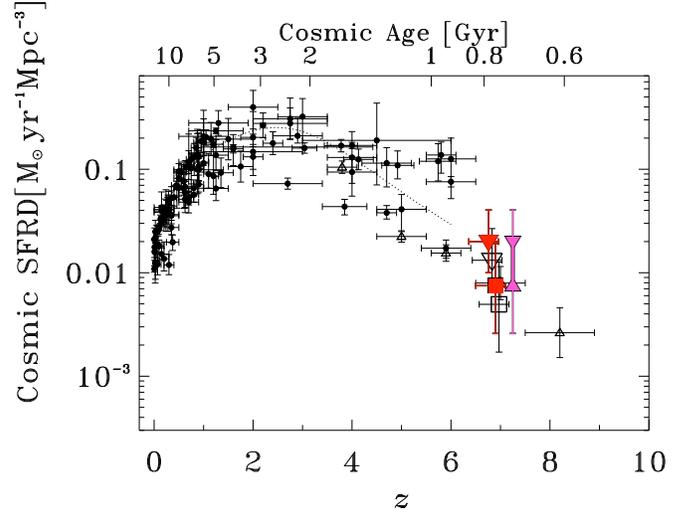}
\caption{
Cosmic star-formation rate density (SFRD) as a function
of redshift. Red square and inverse-triangle represent
the extinction corrected SFRDs integrated down to
$L\simeq 0.1 L^*$ (${\rm SFRD}^{\rm obs}_{\rm corr}$),
and to $L=0$ (${\rm SFRD}^{\rm upper}_{\rm corr}$), respectively.
Open square and inverse-triangle are the same, but
with no extinction correction, i.e., ${\rm SFRD}^{\rm obs}$
and ${\rm SFRD}^{\rm upper}$. Magenta line with triangles
give the allowed SFRDs at $z=7$ which are defined by
${\rm SFRD}^{\rm obs}$ and ${\rm SFRD}^{\rm upper}_{\rm corr}$
with associated errors. We shift the magenta line with the triangles
along the abscissa for the presentation purpose.
Filled circles indicate total SFRDs
at $z\lesssim 6$ that are compiled by \citet{hopkins2006}.
Dotted line denotes the best-fit function to 
the SFRD measurements at $z=0-6$ with the \citet{cole2001} formalism 
\citep{hopkins2006}.
Open triangles show SFRD measurements down to $0.06 L^*(z=3)$ with extinction correction,
which are recently reported by \citet{bouwens2009b}. Because their $z=7$ measurement
is very close to our results, their triangle symbol at $z=7$ is hidden behind 
the red filled square.
\label{fig:sfrd_zdrop}}
\end{figure}

Figure \ref{fig:sfrd_zdrop} plots the cosmic SFRDs from our measurements
(squares and inverse-triangles)
as well as our upper and lower limits including the $1\sigma$ errors 
(magenta line). Figure \ref{fig:sfrd_zdrop} also 
displays the cosmic SFRDs obtained from previous studies
with the assumption of Salpeter IMF.
At $z\simeq 0-6$, we show the cosmic SFRD measurements
compiled by \citet{hopkins2006}. The compilation of 
\citet{hopkins2006} covers most of SFRD measurements 
made, to date, in various wavelength
including H$\alpha$ \citep{glazebrook1999,tresse2002,hanish2006}, 
mid-infrared \citep{flores1999,perez-gonzalez2005}, 
submm \citep{barger2000,hughes1998}, 
radio \citep{condon2002,serjeant2002}, and
X-ray \citep{georgakakis2003}. 
It also includes results of \citet{giavalisco2004b}, 
\citet{bunker2004}, and \citet{ouchi2004} for $z>4$ SFRDs
estimated from UV luminosities.
Figure \ref{fig:sfrd_zdrop} presents 
the best-fit function to the $z\lesssim 6$ SFRD
measurements with the \citet{cole2001} formalism \citep{hopkins2006}.
We also plot SFRDs at high redshifts recently 
reported by \citet{bouwens2008}. 
The SFRDs of \citet{bouwens2008} are applied for extinction correction,
but integrated down to only $0.2 L^*(z=3)$. Their measurements are 
interpreted as lower limits of total SFRDs 
that are counterparts of our ${\rm SFRD}^{\rm obs}_{\rm corr}$ measurement.
Our measurement is consistent with that of \citet{bouwens2008}.
In Figure \ref{fig:sfrd_zdrop}, 
comparing the \citeauthor{hopkins2006}'s (\citeyear{hopkins2006})
best-fit model function (dotted line) with our constraints of 
upper and lower limits (the magenta line with the error bar),
we find that the cosmic SFRD drops 
from the peak at $z=2-3$ to $z=7$ roughly 
by a factor of $\sim 10$ (at least by a factor of $\gtrsim 6$), but
that the drop is not as large as a factor of $\sim 100$.
Figure \ref{fig:sfrd_zdrop} indicates that the cosmic SFRD declines 
even from $z=6$ to $7$, which is originated from the decrease of UV LF
from $z=6$ to $7$. 
Note that this decline of SFRD could be
weaker, if there exist a large population of
very faint galaxies, such suggested by \citet{stark2007}
for $z=9-10$ galaxies, 
that the present blank field surveys cannot identify.
However, the decreasing tendency of SFRD
from $z=2-3$ to $7$ cannot be rejected, 
if the faint-end LF has
a reasonable slope, $\alpha \gtrsim -1.97$,
that changes $\rho_{\rm UV}^{\rm upper}$, i.e., 
${\rm SFRD}^{\rm upper}_{\rm corr}$ 
by a factor of $\lesssim 10$ 
(see \S \ref{sec:uv_luminosity_density}).
Because this suppression of
SFRD towards high redshifts follows 
the evolutionary tendency of dark halo mass function
such as \citet{sheth1999}, this would suggest 
that we would be witnessing the early phase of
galaxy formation history at $z=7$.

\begin{figure}
\epsscale{1.2}
\plotone{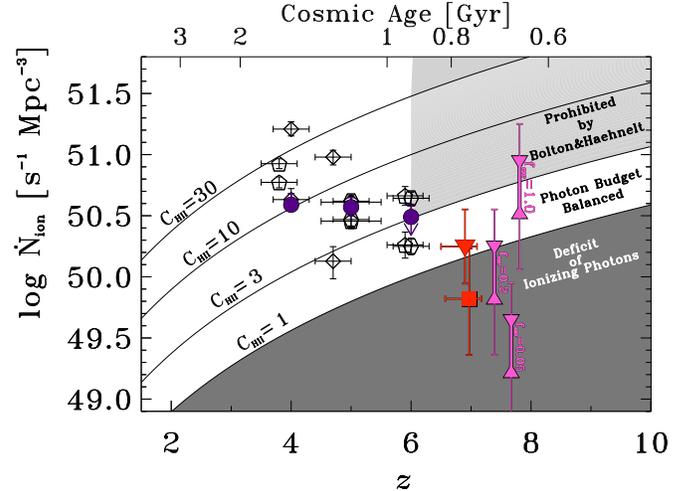}
\caption{
Emission rate of ionizing photon per comoving
Mpc$^3$, $\dot{N}_{\rm ion}$, as a function of redshift.
We assume $f_{\rm esc}=0.2$, if not 
otherwise specified. Square and triangle present 
the lower and upper limits of
$\dot{N}_{\rm ion}$ at $z\simeq 7$ estimated from 
$\rho_{\rm UV}^{\rm obs}$ and $\rho_{\rm UV}^{\rm upper}$,
respectively. Three thick magenta lines with triangle/inverse-triangle
represent the allowed $\dot{N}_{\rm ion}$ ranges at $z=7$ for
$f_{\rm esc}=0.2$, $0.05$, and $1.0$ from left to right.
These ranges are defined by the lower and upper limits
of $\dot{N}_{\rm ion}$. Thin magenta lines denote 
the associated $1\sigma$ errors corresponding 
to $1\sigma$ errors of the upper and lower limit estimates.
For the presentation purpose, we shift the red square and
the magenta lines along the abscissa. The exact redshift is
the same as the one of red triangle.
Solid lines plot the model predictions of $\dot{N}_{\rm ion}$ that
is required for maintaining hydrogen ionization in IGM
\citep{madau1999} with clumping factors of $C_{\rm HII}=1$, $3$,
$10$, and $30$, from bottom to top. Dark gray area
indicates that even the homogeneous Universe ($C_{\rm HII}=1$) 
lacks ionizing photons to maintain hydrogen ionization of IGM
in the model of \citet{madau1999}. 
The photoionization rates inferred from the Ly$\alpha$ forest
are shown with purple circles 
\citep{bolton2007}.
Light gray shade displays constraints from \citet{bolton2007}
who find that the clumping factor is $C_{\rm HII}\lesssim 3$
at $z\sim 6$ (and at $z\gtrsim 6$ in the hierarchical Universe).
Diamonds and pentagons are $\dot{N}_{\rm ion}$ at $z=4-6$
produced by both AGN and galaxies that are calculated 
from AGN LFs \citep{bolton2007} and galaxy LFs given 
by \citet{yoshida2006} and \citet{bouwens2008},
respectively. Upper and lower sequences of diamonds/pentagons
present the upper and lower limits of $\dot{N}_{\rm ion}$ calculated from
their LFs down to the observation limiting luminosities and $L=0$, respectively,
which are counterparts of our $\rho_{\rm UV}^{\rm obs}$ and $\rho_{\rm UV}^{\rm upper}$
measurements. Although our $\dot{N}_{\rm ion}$ at $z=7$ includes no AGN contribution
due to no AGN LF measurements at $z=7$, the contribution of AGN is probably smaller
than $\lesssim 0.08$ dex at $\log \dot{N}_{\rm ion}\simeq 50$ (see text).
\label{fig:ionization_photon}}
\end{figure}

\subsection{Ionization Photon Budget Near the Reionization Epoch}
\label{sec:reionization}

We evaluate emission rate of hydrogen ionizing photon per comoving
Mpc$^3$, $\dot{N}_{\rm ion}$, and discuss 
ionizing photon budget, i.e., whether
the photon production rate of galaxies is
larger than the recombination rate of hydrogen IGM.
We calculate 
$\dot{N}_{\rm ion}$ for galaxies with 
\begin{equation}
\dot{N}_{\rm ion} ({\rm s}^{-1}\ {\rm Mpc}^{-3}) = 10^{49.7} \left( \frac{\epsilon^{\rm g}}{10^{25}} \right) \left( \frac{\alpha_{\rm s}}{3} \right)^{-1} \left( \frac{f_{\rm esc}}{0.1} \right),
\label{eq:ionizingphoton_galaxy}
\end{equation}
where $\epsilon^{\rm g}$ is the ionizing emission density at the Lyman limit
in units of ${\rm erg}\ {\rm s}^{-1} {\rm Hz}^{-1} {\rm Mpc}^{-3}$, 
$\alpha_{\rm s}$ is 
the spectral index of ionizing emission, and $f_{\rm esc}$ is the
escape fraction of ionizing photons \citep{bolton2007}.
We adopt $\epsilon^{\rm g}=\rho_{\rm UV}/6$ for the spectral break
at the Lyman limit, where $\rho_{\rm UV}$
is the UV luminosity density at $\sim 1500$\AA\ \citep{madau1999},
and assume $\alpha_{\rm s}=3$ that corresponds to
a model spectrum of \citet{leitherer1999} with continuous
star-formation history, Salpeter IMF, and a metallicity of 
$Z=0.2Z_\odot$ \citep{bolton2007}. We apply three $f_{\rm esc}$
values; $f_{\rm esc}\simeq 0.2$ suggested by
the recent numerical simulations of \citet{yajima2009},
$f_{\rm esc}\simeq 0.05$ found in LBGs at $z\sim 3$ 
(\citealt{shapley2006}; see also \citeauthor{iwata2009}'s \citeyear{iwata2009}
subsequent studies), and $f_{\rm esc}=1.0$ 
for the maximum escape fraction.
Note that $f_{\rm esc}\simeq 0.05$ is also suggested for $z>4$ galaxies 
in the semi-analytical model of \citet{wyithe2009}.
Figure \ref{fig:ionization_photon} presents two $\dot{N}_{\rm ion}$ values
at $z=7$ estimated from our $\rho_{\rm UV}^{\rm obs}$ and 
$\rho_{\rm UV}^{\rm upper}$ with $f_{\rm esc}=0.2$.
We regard these two $\dot{N}_{\rm ion}$
as the lower and upper limits (see \S \ref{sec:uv_luminosity_density}).
To evaluate the impacts of escape fraction determination,
we plot three magenta lines showing the lower and upper limits
of $\dot{N}_{\rm ion}$ with the three different escape fractions,
$f_{\rm esc}=0.2$, $0.05$, and $1.0$. We calculate
$\dot{N}_{\rm ion}$ at $z=4-6$ from the best-fit
Schechter parameters of UV LFs obtained by 
\citet{yoshida2006}, \citet{bouwens2008}, and \citet{mclure2009a}.
Evaluating upper and lower limits of $\dot{N}_{\rm ion}$ at $z=4-6$,
we integrate UV LFs down to $L=0$ for the upper limits
and down to the observed magnitudes of $M_{\rm UV}\simeq -18$
($z=6$), $M_{\rm UV}\simeq -17$ ($z=5$), and 
 $M_{\rm UV}\simeq -16$ ($z=4$) for the lower limits 
of \citet{bouwens2008} and \citet{mclure2009a}
\footnote{
The measurements of \citet{mclure2009a} 
include only those at $z=6$ and $5$.
}.
Similarly, we integrate UV LFs down to 
$M_{\rm UV}=-20.3$ and $-19.2$
for the lower limits of \citet{yoshida2006} 
at $z=5$ and $4$, respectively.
These UV luminosity densities are used to estimate
$\dot{N}_{\rm ion}$ with eq. (\ref{eq:ionizingphoton_galaxy}) and 
$f_{\rm esc}=0.2$.
Then we add ionizing photons from AGN given by
\citet{bolton2007}, and plot them
in Figure \ref{fig:ionization_photon}. We find that 
$\dot{N}_{\rm ion}$ decreases monotonically from $z=4$ to $7$.

Figure \ref{fig:ionization_photon} also shows $\dot{N}_{\rm ion}$ 
that is required to balance recombination of hydrogen IGM 
based on the model of \citet{madau1999},
\begin{equation}
\dot{N}_{\rm ion} ({\rm s}^{-1}\ {\rm Mpc}^{-3}) =10^{47.4} C_{\rm HII} (1+z)^3
\label{eq:IGMrecombination}
\end{equation}
with clumping factors of $C_{\rm HII}=1$, $3$, $10$, and $30$.
Note that $C_{\rm HII}=1$ corresponds to
the homogeneous Universe, and that the Universe at $z\sim 7$ 
should have $C_{\rm HII}>1$ due to the hierarchical structure formation.
If the photon production rates of objects fall below
the model of $C_{\rm HII}=1$, hydrogen IGM cannot maintain
the ionized state for any clumping factors taken at the redshift.
On the other hand, the photoionization rates
inferred from the Ly$\alpha$ forest indicate that 
clumping factor is as small as $C_{\rm HII}\lesssim 3$ at $z\sim 6$
(\citealt{bolton2007}; the purple circles in Figure \ref{fig:ionization_photon}). 
Since $C_{\rm HII}$ should monotonically decrease
towards high redshifts in the hierarchical Universe, 
the models with $C_{\rm HII}\lesssim 3$  can be applied to
the Universe at $z\gtrsim 6$. 
If $\dot{N}_{\rm ion}$ of objects falls in or beyond the model of 
$1<C_{\rm HII}\lesssim 3$ at $z\gtrsim 6$, the ionizing photon production rate 
is high enough to maintain the ionized IGM.

Figure \ref{fig:ionization_photon} presents 
that, in the cases of $f_{\rm esc}=0.2$ and 
$f_{\rm esc}=1.0$,
$\dot{N}_{\rm ion}$ values of $z=7$ galaxies 
are comparable with those predicted by
the models of $1<C_{\rm HII}\lesssim 3$.
We do not include the AGN contribution of $\dot{N}_{\rm ion}$ 
to our estimate at $z=7$, because no AGN UV LF data are available at this redshift.
However, AGN contribution of ionizing photon production is 
only $\log \dot{N}_{\rm ion}\simeq 49.3$ at $z=6$ \citep{bolton2007}.
Because the comoving density of luminous QSOs at $z\sim 6$ 
is 30 times smaller than that at $z\sim 3$ \citep{fan2004},
AGN LF probably continues decreasing towards $z\sim 7$.
Even if we assume no evolution of AGN LF from $z=6$ to $7$,
the AGN contribution is negligible; only pushing 0.08 dex at 
$\log \dot{N}_{\rm ion}\simeq 50$ 
in Figure \ref{fig:ionization_photon}.

Our results indicate that the ionizing photon budget
just balances at $z=7$ for $f_{\rm esc}\gtrsim 0.2$.
On the other hand, $\dot{N}_{\rm ion}$ of $z=7$ galaxies
with $f_{\rm esc}=0.05$ is a factor of three below 
the model of $C_{\rm HII}=1$, which has 
the $\simeq 95$\% ($2\sigma$) confidence level.
Note that $f_{\rm esc}\simeq 0.05$ is 
the measured escape fraction at $z\sim 3$ \citep{shapley2006}. 
Moreover, all of the other parameters for our $\dot{N}_{\rm ion}$ estimates, 
i.e. the spectral index and break, 
are plausible ones for low-$z$ star-forming galaxies.
Thus, there are two scenarios. a) If no properties of star-forming galaxies at 
$z=7$ are different from those at low redshifts, 
the universe could not be totally ionized by only galaxies (and AGN)
at $z=7$ at the $\simeq 95$\% confidence level.
b) If the properties of
star-forming galaxies evolve from low redshifts, 
e.g. larger $f_{\rm esc}$ ($f_{\rm esc}\gtrsim 0.2$),
lower metallicity, flatter IMF, 
and/or less dust extinction, the Universe at $z=7$ 
is ionized and close to being in balance between the rates of 
ionizing photon production and recombination of hydrogen IGM.
If the scenario a) is true, 
the hydrogen IGM would experience a deficit of 
ionizing photon at $z=7$.
This implies that the Universe may not complete
the reionization by $z=7$. In this case, 
the Universe would start reionization right after $z=7$ 
and almost complete it by $z\sim 6$ \citep{fan2006}.
However, the polarization data of WMAP5 place the constraints that
instantaneous reionization below $z=8.2$ ($=6.7$) is rejected at
the $2\sigma$ ($3\sigma$) level \citep{dunkley2009}. It is unlikely that 
reionization of the Universe is completed 
at the late epoch of $z\sim 6-7$ almost instantaneously.
Alternatively, the scenario of a) indicates that 
not galaxies and AGN but some exotic populations or
physical phenomena might play an important role 
in cosmic reionization.
Although several candidates of the exotic ionizing sources
are suggested, such as dark matter annihilation
and primordial black holes (e.g. \citealt{ricotti2004,ricotti2008}),
no observational evidence have been obtained.
It is more likely that the scenario b) is correct, 
and that the escape fraction, metallicity, IMF, and dust extinction
are larger, lower, flatter, and/or smaller, respectively.
In fact, \citet{inoue2006} claim that $f_{\rm esc}$ increases
towards high redshifts at $0<z<6$
via their model calculations.
The numerical simulations of \citet{razoumov2009} also 
suggest the similar evolution of $f_{\rm esc}$, which
reaches 0.8 at $z=10$, and that the angular averaged escape fraction of 
$f_{\rm esc}=0.3-0.6$ at $z\sim 7$ for galaxies similar to ours
(SFR$\simeq 1-10$ $M_\odot$).
Metallicity of galaxies also impact on the
production of ionizing photons. \citet{stiavelli2004}
claim that the ionizing efficiency of a stellar population 
increases by a factor of 3 for Salpeter IMF and 
a factor of 10 for a top-heavy IMF 
as the metallicity decreases 
from $Z=Z_\odot$ to $Z=0$ (see also \citealt{schaerer2003}).
\footnote{
Since the temperature of IGM increases
from $10,000$ K (for the solar metallicity)
to $\sim 20,000$ K (for low metallicity; \citealt{osterbrock1989}),
the low metallicity in IGM would reduce 
the recombination rate of IGM with the solar metallicity
by a factor of $\sim 2$ \citep{stiavelli2004}.
}
Moreover, dust absorption may be important
in determination of escape fraction, as demonstrated by
the simulations of \citet{laursen2009}. Because Ly$\alpha$ emissivity
would be higher at $z\sim 6$ than low-$z$
as suggested by LAE studies
(e.g. \citealt{ouchi2008}),
a flatter IMF as well as lower metallicity and/or 
less dust extinction
towards high-$z$ may be plausible. 
It should be noted that,
even in the scenario b),
our observational constraints are close to being 
in balance between ionizing photon production
and recombination rates at $z\sim 7$. In other words,
we might be witnessing the final stage of reionization
with the closely balanced photon budget. It would provide
signatures of the neutral fraction evolution 
that are claimed by \citet{iye2006} and \citet{ota2008} who
find significantly less number of LAEs
at $z\sim 7$ than $z\sim 6$. Moreover, the scenario of b) is
very consistent with the extended ($z\sim 6-11$) reionization
picture suggested by \citet{dunkley2009}.
Note that these arguments assume that
there is no emergence of a large population
of very faint galaxies at $z=7$ beyond the
Schechter function, such
claimed by \citet{stark2007} for 
the earlier epoch of $z=9-10$.
On the other hand, \citet{santos2004} have 
found that star-formation activities of 
low-mass galaxies are suppressed 
at $z\simeq 5$
via their Keck gravitational lensing survey.
Systematic lensing surveys for $z=7$ 
galaxies would be needed to correctly 
understand the contribution 
from these very faint galaxies.
Although there remain the arguments
of the very faint galaxy population, 
the $\dot{N}_{\rm ion}$ of $z=7$ galaxies
with $f_{\rm esc}=0.05$ still falls below
the model of $C_{\rm HII}=1$
with a very steep faint-end slope down to 
$\alpha \simeq -1.90$ that would boosts 
$\rho_{\rm UV}^{\rm upper}$, i.e., 
the upper limit of $\dot{N}_{\rm ion}$
by a factor of $\simeq 3$
(\S \ref{sec:uv_luminosity_density}).

\begin{figure}
\epsscale{2.3}
\plottwo{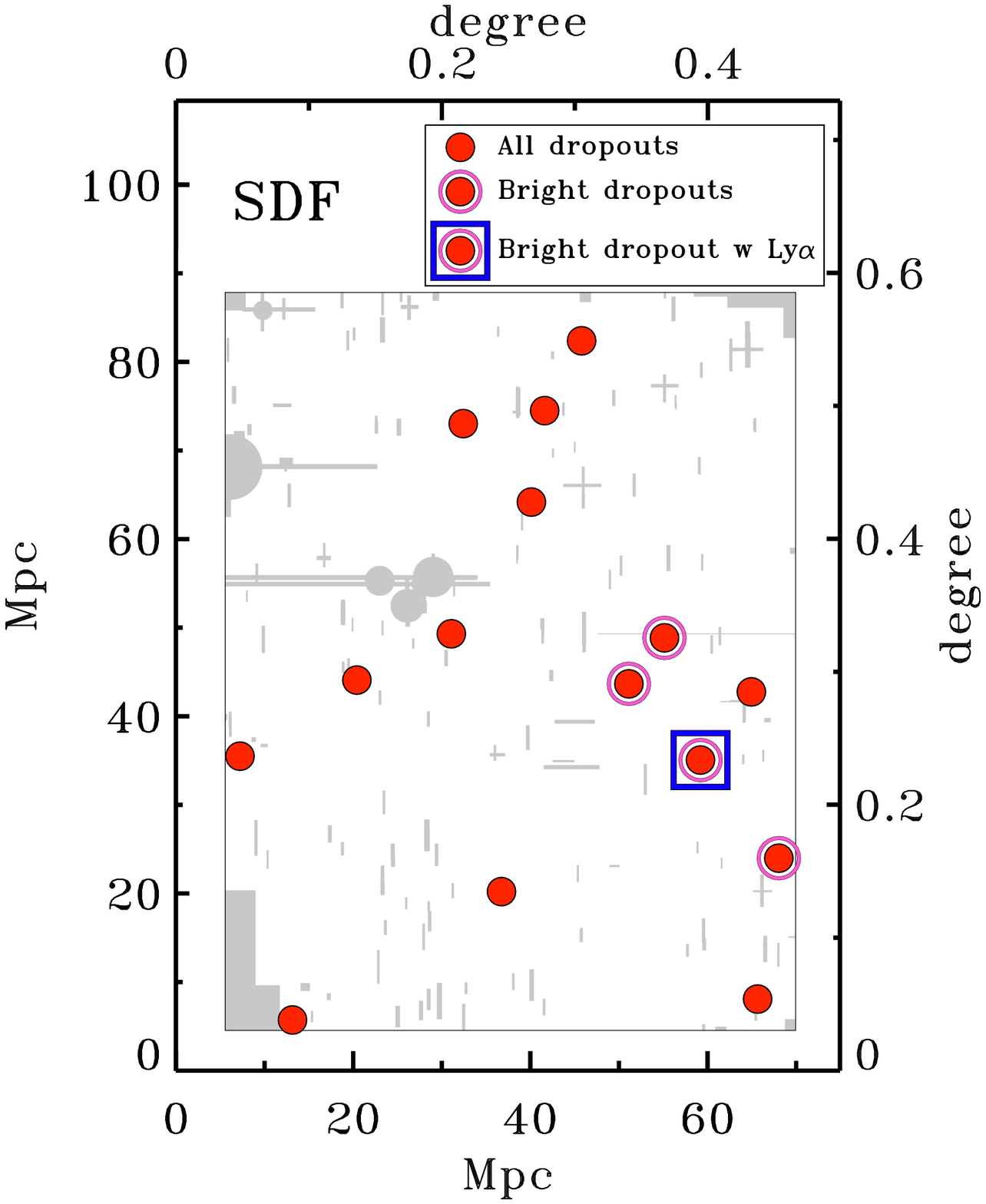}{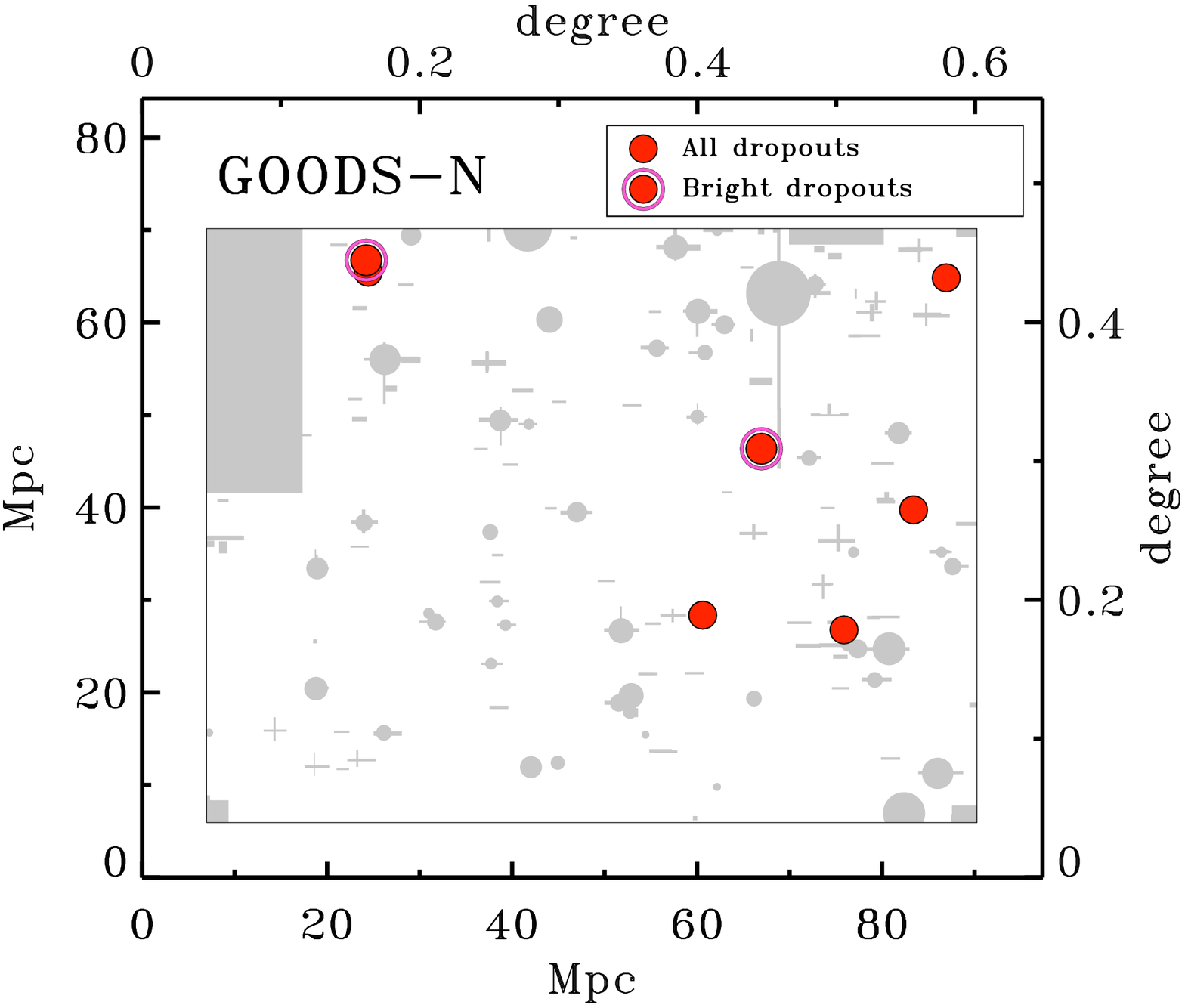}
\caption{
Sky distribution of our $z=7$ dropout galaxies down to $y=26.1$
in SDF (left) and down to $y=25.9$ in GOODS-N (right).
Red circles represent positions of all dropout galaxies,
while red circles with a magenta circle denote bright dropouts
with $y<25.6$. Blue square indicates the position of the bright dropout with
Ly$\alpha$ emission at $z_{\rm spec}=6.96$. Gray shades are
masked areas where we did not use the data for our analysis.
The scales on the maps are marked in both degrees and
comoving megaparsecs in projection at $z=7$.
North is up and east is to the left in these images.
\label{fig:dist_zdrop}}
\end{figure}

\subsection{Distribution of Dropouts: Indication of Ionized Bubble?}
\label{sec:distribution}
Figure \ref{fig:dist_zdrop} presents the
sky distribution of our $z$-dropout galaxies
in SDF and GOODS-N. Although the numbers of galaxies 
in each field are small, they appear to be clustered on
the sky. The distribution in SDF shows possible three 
filamentary structures crossing at around 
the center of the field; from top to the center, the center to 
bottom right, and the center to bottom left. 
The possible filaments would extend up to $\sim 60$ Mpc.
Our dropout galaxies
in GOODS-N are located mostly at the center right and top left.
These sky distributions imply that our $z=7$ galaxies with a bright
UV luminosity ($M_{\rm UV}\sim -21$; $SFR_{\rm no dust}=10-30$)
would be strongly clustered, which are similar to
those at $z\sim 6$ \citep{ouchi2005,overzier2006,lee2009}.
In Figure \ref{fig:dist_zdrop},
we mark the positions of
the UV brightest galaxies with $y<25.6$
among our galaxies at $z=7$.
We find that these brightest galaxies are located at
the high density regions of $z$-dropout
galaxies both in SDF and GOODS-N.
In the SDF panel of Figure \ref{fig:dist_zdrop}, 
we also mark the Ly$\alpha$ emitting
dropout galaxy, SDF-63544, at $z_{\rm spec}=6.96$ 
confirmed by spectroscopy \citep{iye2006}.
Interestingly, this Ly$\alpha$ emitting
dropout galaxy sits at the center of
the 4 UV brightest dropout galaxies
whose distribution extends by $\sim 30$ Mpc
in projection (Figure \ref{fig:dist_zdrop}).
Because \citet{ota2008} did not confirm
the other Ly$\alpha$ emitter candidate found by 
their narrow-band survey of $z=7$ 
even with their deep spectroscopic data, 
this SDF-63544 may be only one 
with an observable Ly$\alpha$ emission line at $z=7$ in the SDF.
It would be possible that an overdense region of the 4 UV brightest
dropouts would make a well-established ionized bubble
of IGM in the cosmic volume with a size of $\gtrsim 30$ Mpc, 
and that the ionized bubble may allow SDF-63544
to transmit Ly$\alpha$ to observers
with no strong Ly$\alpha$ damping absorption 
given by neutral hydrogen of IGM.
To evaluate how much Ly$\alpha$ flux 
is absorbed by IGM, we estimate a Ly$\alpha$ equivalent width of 
the Ly$\alpha$ emitting dropout galaxy
from the UV continuum magnitude ($-21.40 \pm 0.31$)
and Ly$\alpha$ flux ($2\times 10^{-17}$ erg s$^{-1}$ cm$^{-2}$; 
see \S \ref{sec:uv_luminosity_function}), 
and obtain the rest-frame equivalent width of 
$EW_0=37\pm 8$\AA. This $EW_0$ is about a half of
the one for the case B recombination 
with no absorption of Ly$\alpha$
(68\AA; \citealt{nagamine2008}; Ono Y. et al. 2009
submitted to MNRAS). We define the escape fraction
of Ly$\alpha$ emission, $f_{\rm esc}^{\rm Ly\alpha}$, by
\begin{equation}
f_{\rm esc}^{\rm Ly\alpha}=\frac{L_{\rm obs}^{\rm Ly\alpha}}{L_{\rm int}^{\rm Ly\alpha}},
\label{eq:lya_escape_fraction}
\end{equation}
where $L_{\rm obs}^{\rm Ly\alpha}$ and $L_{\rm int}^{\rm Ly\alpha}$
are observed and intrinsic Ly$\alpha$ luminosities, respectively.
If we assume the case B recombination, 
$L_{\rm int}^{\rm Ly\alpha}[{\rm erg s}^{-1}]=1.1\times 10^{42}$ SFR[$M_\odot$yr$^{-1}$],
and SFR from $M_{\rm UV}$ with no dust extinction correction (eq. \ref{eq:uv_sfr})
\footnote{
SFR thus obtained is 19.7 $M_\odot$ yr$^{-1}$ 
for the Ly$\alpha$ emitting dropout.
},
we estimate the Ly$\alpha$ escape fraction of the Ly$\alpha$ emitting
dropout to be $f_{\rm esc}^{\rm Ly\alpha}=0.54 \pm 0.12$.
This $f_{\rm esc}^{\rm Ly\alpha}$ would be comparable to
the one at $z\sim 6$ with no strong IGM damping wing of 
Ly$\alpha$, because the average Ly$\alpha$ escape fraction
is $f_{\rm esc}^{\rm Ly\alpha}=0.52-0.54$ 
at $z\sim 6$ estimated from the Ly$\alpha$ opacities of 
\citet{fan2006}, \citet{madau1995}, and \citet{meiksin2006} 
for the case that IGM absorbs
a blue half of symmetric Ly$\alpha$ emission line (see \S 5.1.2 of
\citealt{ouchi2008}).
The inferred $f_{\rm esc}^{\rm Ly\alpha}$
would support the idea
that the Ly$\alpha$ emitting dropout galaxy 
sits inside a well-established ionized bubble with
a neutral fraction as low as that at $z\sim 6$.

If this naive physical picture is correct, the size of ionized bubble
would be larger than $30$ Mpc at $z=7$. Assuming that this
is a lower limit of the characteristic bubble size,
we find that the analytic models of \citet{furlanetto2006}
would suggest an upper limit of neutral fraction
of $x_{\rm HI}\lesssim 20$\% at $z=7$.
This small upper limit of $x_{\rm HI}$ may 
indicate that the Universe is not fully neutral at $z=7$.
The combination of reionization models and 
LAE LF (+clustering) gives constraints
on neutral fraction of $x_{\rm HI}\lesssim 50$\% at $z=6.5$
(\citealt{dijkstra2007,mcquinn2007,iliev2008}) 
and $x_{\rm HI}\lesssim 10-60$\% at $z=7$ \citep{kobayashi2007,ota2008}. 
Our possible upper limit of $x_{\rm HI}\lesssim 20$\% at $z=7$
is consistent with those of previous results
from the independent observational probes.
This implies that there is no strong evidence rejecting 
the presence of the $\gtrsim 30$-Mpc ionized bubble
at $z=7$.

\section{Conclusions}
\label{sec:conclusions}

We have identified 22 $z$-dropout galaxy candidates
in the 0.4 deg$^2$ area of SDF and GOODS-N down to $y=26$ 
with deep $\simeq 30$-hour Subaru $y$-band images and 
Subaru and HST legacy imaging data. One out of 22 $z$-dropout 
galaxies in the SDF has a spectroscopic redshift of $z=6.96$ determined
from Ly$\alpha$ emission. We have derived the bright-end UV LF of 
galaxies at $z=7$. Based on our bright $z$-dropout galaxies
as well as faint $z$-dropout galaxies obtained by the recent 
HST/WFC3 studies, we have constrained the early stage of galaxy formation
and photon budget of cosmic reionization at $z=7$.
We have also discussed 
the distributions of our $z$-dropout galaxies.
The major results of our study are summarized below.

1. We find that our bright-end UV LF shows a decrease 
from $z\sim 6$ to $7$. This decrease cannot be explained
by statistical errors or cosmic variance
that is less than a factor of $\lesssim 2$ in number density
for a 0.2 deg$^2$ area.
The best-fit Schechter parameters of $z=7$ galaxies
are 
$\phi^*=0.69_{-0.55}^{+2.62} \times 10^{-3}$Mpc$^{-3}$,
$M_{\rm UV}^* = -20.10\pm 0.76$ mag, and
$\alpha = -1.72\pm 0.65$.
Our Schechter parameter fit results reject
no evolution of UV LF from $z=6$ to $7$
at the $>95$\% confidence level.
A more dominant decrease of $L^*$ than $\phi^*$
is preferable from $z=5-6$ to $7$ 
in the error contours of Schechter parameters.

2. The cosmic SFRD drops from $z=2-3$ to $z=7$ by a factor of 
at least $\gtrsim 6$. It is likely that the cosmic SFRD decreases
roughly by a factor of $\sim 10$, but not larger than $\sim 100$. This suppression of
cosmic SFRD implies that we would be witnessing the early phase of
galaxy formation history.

3. If we assume the properties of low-$z$ star-forming galaxies
including the spectral shape and escape fraction ($f_{\rm esc}\simeq 0.05$),
we find that the ionized photon production rate of $z=7$ galaxies
falls below the hydrogen IGM recombination rate 
predicted by the analytic models of \citet{madau1999}
even in the homogeneous Universe ($C_{\rm HII}=1$)
at the $\simeq 95$\% ($2\sigma$) confidence level.
Although it implies that 
the Universe cannot be totally ionized by only galaxies
at $z=7$,
but we think that properties of galaxies 
at $z=7$ are just different from those at low redshifts with, 
e.g., a larger escape fraction ($f_{\rm esc}\gtrsim 0.2$), 
a lower metallicity, a flatter IMF, and/or less dust extinction.
In either case, our observational constraints imply that
the Universe at $z\sim 7$ is close to being in balance between
rates of ionizing photon production and IGM hydrogen recombination.

4. $z$-dropout galaxies may be strongly clustered
both in SDF and GOODS-N. We find that the distribution
of $z$-dropout galaxies
in SDF appears to be a filamentary shape which extends 
up to $60$ Mpc in projection, and that the $z=6.96$ dropout galaxy 
with a Ly$\alpha$ line is located at the center of the overdense region 
consisting of the 4 UV brightest dropout galaxy candidates. 
This implies that there may exist a well-established 
ionized bubble made by the 4 UV brightest dropout galaxies,
and that the ionized bubble might help to transmit 
the Ly$\alpha$ line in IGM at $z=7$.

\acknowledgments
We thank Ross Mclure, Rychard Bouwens, Pascal Oesch,
and Andrew Hopkins
for providing their data. 
We are grateful to
Daniel Schaerer, Wei Zheng, Min-Su Shin, and David Sobral 
for their useful comments.
We acknowledge the current and former Subaru Observatory staff,
especially Hisanori Furusawa, Akito Tajitsu, Miki Ishii,
Michihiro Takami, and Fumiaki Nakata, for their invaluable help 
that made this challenging and long-standing project possible.
M.O. has been supported via Carnegie Fellowship.

{\it Facilities:} \facility{Subaru (Suprime-Cam)}

\clearpage

\begin{deluxetable}{cccrcccl}
\tablecolumns{8}
\tabletypesize{\scriptsize}
\tablecaption{Summary of Imaging Observations and Data 
\label{tab:obs}}
\tablewidth{0pt}
\tablehead{
\colhead{Field} & 
\colhead{Band} & 
\colhead{CCD} &
\colhead{t$_{\rm exp}$} &
\colhead{PSF size\tablenotemark{a}} &
\colhead{Area} &
\colhead{$m_{\rm lim}$\tablenotemark{b}}  &
\colhead{Date of Observations and Note} \\
\colhead{} & 
\colhead{} & 
\colhead{} & 
\colhead{(sec)} &
\colhead{(arcsec)} &
\colhead{(arcmin$^2$)} &
\colhead{(3$\sigma$ AB mag)}  &
\colhead{}
}
\startdata
SDF     & $y$   & MIT-LL &  50614 & \nodata & \nodata & \nodata & 2007 Apr. 17-20, (2003 Jul. 2, 2004 Mar. 19-20)\tablenotemark{c}\\
        & $y$   & Hamamatsu & 43901 & \nodata & \nodata & \nodata & 2009 Apr. 24-26\\
        & $y$   & (total) & 94515 & 0.79 (0.99) & 810 & 26.4 & \nodata \\
        & $z'$  & MIT-LL & 104069 & 0.91 (0.99) & 810 & 27.7 & Data from Morokuma et al. in preparation\tablenotemark{d}\\
\hline
GOODS-N & $y$   & MIT-LL &  89308 & \nodata & \nodata & \nodata & 2006 Apr. 1, 3, 2007 Apr. 17-18,20  \\
        & $y$   & Hamamatsu &  29101 & \nodata & \nodata & \nodata & 2009 Apr. 24-25, 27  \\
        & $y$   & (total) &  118409 & 0.87 (0.87) & 758 & 26.2 & \nodata  \\
        & $z'$ & MIT-LL &  39150 & 0.87 (0.87) & 758 & 26.9 & 2007 Apr. 19, (2001 Apr. 20, 22, \\
        &      &        &        &             &       &      & 2003 Apr. 6, 2004 Mar. 16, 2006 Feb. 23-24)\tablenotemark{c}
\enddata
\tablenotetext{a}{
The FWHM of PSFs in the reduced image. The values in parenthesis
indicate the FWHM of PSFs that are matched with broad-band images 
in each field.
}
\tablenotetext{b}{
The limiting magnitude defined by 
a $3\sigma$ sky noise level 
in a $1''.8$-diameter circular aperture.
}
\tablenotetext{c}{
The observation dates in parenthesis are those of 
the Subaru archival data taken by the other teams.
}
\tablenotetext{d}{
See \S \ref{sec:data}.
}
\end{deluxetable}

\clearpage

\begin{deluxetable}{ccc}
\tabletypesize{\scriptsize}
\tablecaption{Limiting Magnitudes
\label{tab:limitingmag}}
\tablewidth{0pt}
\tablehead{
\colhead{Band} &
\colhead{GOODS-N} &
\colhead{SDF\ \ \ }\\
\colhead{} &
\colhead{(1)} &
\colhead{(2)}
}
\startdata
$U$                   & $26.9$\tablenotemark{1} & \nodata \\
$B$                   & $27.5$                  & $28.9$  \\
$B_{435}$             & $28.4$                     & \nodata \\
$V$                   & $26.2$\tablenotemark{1} & $28.1$  \\
$V_{606}$             & $28.6$                     & \nodata \\
$R$                   & $26.5$\tablenotemark{1} & $28.4$  \\
$I$ or $i'$           & $26.5$                  & $28.1$  \\
$i_{775}$             & $27.9$                     & \nodata    \\
$z_{850}$             & $27.6$                     & \nodata    \\
$z'$\tablenotemark{2} & $26.9$                  & $27.7$  \\
$y$\tablenotemark{2}  & $26.2$                  & $26.4$  \\
$m_{3.6}$    & $26.1$                     & \nodata \\
$m_{4.5}$    & $25.5$                     & \nodata \\
$m_{5.8}$    & $23.5$                     & \nodata \\
$m_{8.0}$    & $23.4$                     & \nodata \\
$m_{24}$     & $21.2$                     & \nodata \\
\enddata

\tablecomments{
Three sigma limiting magnitudes 
in GOODS-N (1) and SDF (2).
The magnitudes are defined with a
$1''.8$-diameter aperture for the ground-based
optical images ($UBVRizy$),
except for the GOODS-N $U$, $V$, and $R$, data.
We apply 
a $0''.2$-diameter aperture for the HST images
($B_{435}$, $V_{606}$, $i_{775}$, $z_{850}$),
and
a $3''.0$-diameter for the Spitzer $3.6\mu$m,
$4.5\mu$m, $5.8\mu$m, $8.0\mu$m, and $24\mu$m images
($m_{3.6}$, $m_{4.5}$, $m_{5.8}$, $m_{8.0}$, $m_{24}$).
The limiting magnitudes of Spitzer data include the offsets
of aperture corrections (see text).
}
\tablenotetext{1}{
The detection limits of ground-based GOODS-N $U$, $V$, and $R$ images are 
defined with a $3''.0$-diameter aperture
due to the relatively poor seeing sizes ($FWHM\simeq 1''.5$; see the text).
}
\tablenotetext{2}{
The $4\sigma$ detection limits of $y$ magnitudes are
$25.9$ (GOODS-N) and $26.1$ (SDF).
The $2\sigma$ upper limits of $z'$ magnitudes are
$27.4$ and $28.2$ in GOODS-N and SDF, respectively.
}
\end{deluxetable}

\clearpage

\begin{deluxetable}{ccccccccccccl}
\tablecolumns{13}
\tabletypesize{\scriptsize}
\tablecaption{$z=7$ Galaxy Candidates
\label{tab:sample}}
\setlength{\tabcolsep}{0.02in}
\tablewidth{0pt}
\tablehead{
\colhead{\scriptsize ID} &
\colhead{\scriptsize $U$} &
\colhead{\scriptsize $B$} &
\colhead{\scriptsize $V$} &
\colhead{\scriptsize $R$} &
\colhead{\scriptsize $I/i$} &
\colhead{\scriptsize $z$} &
\colhead{\scriptsize $y$} &
\colhead{\scriptsize $m_{3.6}$} &
\colhead{\scriptsize $m_{4.5}$} &
\colhead{\scriptsize $z'-y$} &
\colhead{\scriptsize $z_{\rm spec}$} &
\colhead{\scriptsize Note}
}
\startdata
\cutinhead{GOODS-N}
{\scriptsize GOODSN152505$^a$} & {\scriptsize$>27.3$} & {\scriptsize$>28.0$} & {\scriptsize$>26.7$} & {\scriptsize$>27.0$} & {\scriptsize$>27.0$} & {\scriptsize$>27.4$} & {\scriptsize$25.50$} & {\scriptsize$>23.4$\tablenotemark{b}} & {\scriptsize$>24.2$\tablenotemark{b}} & {\scriptsize$>1.90$} & \nodata & {\scriptsize Source conf. in Spitzer bands}\\
              &         & \scriptsize ($>28.4$) & \scriptsize ($>28.6$)   &   & \scriptsize ($>27.9$) & \scriptsize ($>27.6$) &    &       &       &      &         & \\
{\scriptsize GOODSN206815} & \scriptsize $>27.3$ & \scriptsize $>28.0$ & \scriptsize $>26.7$ & \scriptsize $>27.0$ & \scriptsize $>27.0$ & \scriptsize $>27.4$ & \scriptsize $25.52$ & \nodata & \nodata & \scriptsize $>1.88$ & \scriptsize \nodata & \nodata \\
{\scriptsize GOODSN201340} & \scriptsize $>27.3$ & \scriptsize $>28.0$ & \scriptsize $>26.7$ & \scriptsize $>27.0$ & \scriptsize $>26.9$\tablenotemark{b} & \scriptsize $>27.4$ & \scriptsize $25.69$ & \nodata & \nodata & \scriptsize $>1.71$ & \nodata & \scriptsize Source confusion\\
{\scriptsize GOODSN104059} & \scriptsize $>27.3$ & \scriptsize $>28.0$ & \scriptsize $>26.7$ & \scriptsize $>27.0$ & \scriptsize $>27.0$ & \scriptsize $>27.4$ & \scriptsize $25.79$ & \scriptsize \nodata & \nodata & \scriptsize $>1.61$ & \nodata & \nodata \\
{\scriptsize GOODSN108036$^a$} & \scriptsize $>27.3$ & \scriptsize $>28.0$ & \scriptsize $>26.7$ & \scriptsize $>27.0$ & \scriptsize $>27.0$ & \scriptsize $>27.4$ & \scriptsize $25.82$ & \scriptsize $>24.8$\tablenotemark{b} & \scriptsize $>24.4$\tablenotemark{b} & \scriptsize $>1.58$ & \nodata & \scriptsize Source conf. in Spitzer bands\\
              &         & \scriptsize ($>28.4$) & \scriptsize ($>28.6$)   & \scriptsize   & \scriptsize ($>27.9$) & \scriptsize ($>27.6$) &    &       &       &      &         & \\
{\scriptsize GOODSN203550} & \scriptsize $>27.3$ & \scriptsize $>28.0$ & \scriptsize $>26.7$ & \scriptsize $>27.0$ & \scriptsize $>27.0$ & \scriptsize $>27.4$ & \scriptsize $25.83$ & \scriptsize \nodata & \scriptsize \nodata & \scriptsize $>1.57$ & \scriptsize \nodata & \scriptsize Spurious source? \\
{\scriptsize GOODSN134896} & \scriptsize $>27.3$ & \scriptsize $>28.0$ & \scriptsize $>26.7$ & \scriptsize $>27.0$ & \scriptsize $>27.0$ & \scriptsize $>27.4$ & \scriptsize $25.86$ & \scriptsize \nodata & \scriptsize \nodata & \scriptsize $>1.54$ & \scriptsize \nodata & \scriptsize \nodata \\
\cutinhead{SDF}
{\scriptsize SDF63544} & \scriptsize \nodata & \scriptsize $>29.4$ & \scriptsize $>28.6$ & \scriptsize $>28.8$ & \scriptsize $>28.6$ & \scriptsize $27.02$ & \scriptsize $25.42$ & \scriptsize \nodata & \scriptsize \nodata & \scriptsize 1.60 & \scriptsize 6.96    &\scriptsize Spectrum in \cite{iye2006} \\
{\scriptsize SDF83878} & \scriptsize \nodata & \scriptsize $>29.4$ & \scriptsize $>28.6$ & \scriptsize $>28.8$ & \scriptsize $>28.6$ & \scriptsize $27.04$ & \scriptsize $25.49$ & \scriptsize \nodata & \scriptsize \nodata & \scriptsize 1.55 & \scriptsize \nodata & \scriptsize \nodata \\
{\scriptsize SDF46975} & \scriptsize \nodata & \scriptsize $>29.4$ & \scriptsize $>28.6$ & \scriptsize $>28.8$ & \scriptsize $>28.6$ & \scriptsize $27.48$ & \scriptsize $25.51$ & \scriptsize \nodata & \scriptsize \nodata & \scriptsize 1.97 & \scriptsize \nodata & \scriptsize \nodata \\
{\scriptsize SDF76507} & \scriptsize \nodata & \scriptsize $>29.4$ & \scriptsize $>28.6$ & \scriptsize $>28.8$ & \scriptsize $>28.6$ & \scriptsize $27.11$ & \scriptsize $25.59$ & \scriptsize \nodata & \scriptsize \nodata & \scriptsize 1.52 & \scriptsize \nodata & \scriptsize \nodata \\
{\scriptsize SDF123919} & \scriptsize \nodata & \scriptsize $>29.4$ & \scriptsize $>28.6$ & \scriptsize $>28.8$ & \scriptsize $>28.6$ & \scriptsize $27.51$ & \scriptsize $25.79$ & \scriptsize \nodata & \scriptsize \nodata & \scriptsize 1.72 & \scriptsize \nodata & \scriptsize \nodata \\
{\scriptsize SDF77202} & \scriptsize \nodata & \scriptsize $>29.4$ & \scriptsize $>28.6$ & \scriptsize $>28.8$ & \scriptsize $>28.6$ & \scriptsize $27.50$ & \scriptsize $25.86$ & \scriptsize \nodata & \scriptsize \nodata & \scriptsize 1.64 & \scriptsize \nodata & \scriptsize \nodata\\
{\scriptsize SDF75298} & \scriptsize \nodata & \scriptsize $>29.4$ & \scriptsize $>28.6$ & \scriptsize $>28.8$ & \scriptsize $>28.6$ & \scriptsize $27.50$ & \scriptsize $25.87$ & \scriptsize \nodata & \scriptsize \nodata & \scriptsize 1.63 & \scriptsize \nodata & \scriptsize \nodata \\
{\scriptsize SDF20911} & \scriptsize \nodata & \scriptsize $>29.4$ & \scriptsize $>28.6$ & \scriptsize $>28.8$ & \scriptsize $>28.6$ & \scriptsize $27.41$ & \scriptsize $25.89$ & \scriptsize \nodata & \scriptsize \nodata & \scriptsize 1.52 & \scriptsize \nodata & \scriptsize \nodata \\
{\scriptsize SDF121488} & \scriptsize \nodata & \scriptsize $>29.4$ & \scriptsize $>28.6$ & \scriptsize $>28.8$ & \scriptsize $>28.6$ & \scriptsize $27.49$ & \scriptsize $25.97$ & \scriptsize \nodata & \scriptsize \nodata & \scriptsize 1.52 & \scriptsize \nodata & \scriptsize \nodata \\
{\scriptsize SDF84539} & \scriptsize \nodata & \scriptsize $>29.4$ & \scriptsize $>28.6$ & \scriptsize $>28.8$ & \scriptsize $>28.6$ & \scriptsize $28.16$ & \scriptsize $25.98$ & \scriptsize \nodata & \scriptsize \nodata & \scriptsize 2.18 & \scriptsize \nodata & \scriptsize \nodata \\
{\scriptsize SDF16416} & \scriptsize \nodata & \scriptsize $>29.4$ & \scriptsize $>28.6$ & \scriptsize $>28.8$ & \scriptsize $>28.6$ & \scriptsize $>28.2$ & \scriptsize $25.99$ & \scriptsize \nodata & \scriptsize \nodata & \scriptsize $>2.21$ & \scriptsize \nodata & \scriptsize \nodata \\
{\scriptsize SDF64206} & \scriptsize \nodata & \scriptsize $>29.4$ & \scriptsize $>28.6$ & \scriptsize $>28.8$ & \scriptsize $>28.6$ & \scriptsize $>28.2$ & \scriptsize $26.07$ & \scriptsize \nodata & \scriptsize \nodata & \scriptsize $>2.13$ & \scriptsize \nodata & \scriptsize \nodata \\
{\scriptsize SDF107344} & \scriptsize \nodata & \scriptsize $>29.4$ & \scriptsize $>28.6$ & \scriptsize $>28.8$ & \scriptsize $>28.6$ & \scriptsize $>28.2$ & \scriptsize $26.08$ & \scriptsize \nodata & \scriptsize \nodata & \scriptsize $>2.12$ & \scriptsize \nodata & \scriptsize \nodata \\
{\scriptsize SDF136726} & \scriptsize \nodata & \scriptsize $>29.4$ & \scriptsize $>28.6$ & \scriptsize $>28.8$ & \scriptsize $>28.6$ & \scriptsize $27.59$ & \scriptsize $26.08$ & \scriptsize \nodata & \scriptsize \nodata & \scriptsize 1.51 & \scriptsize \nodata & \scriptsize \nodata \\
{\scriptsize SDF41484} & \scriptsize \nodata & \scriptsize $>29.4$ & \scriptsize $>28.6$ & \scriptsize $>28.8$ & \scriptsize $>28.2$\tablenotemark{b} & \scriptsize $28.05$ & \scriptsize $26.09$ & \scriptsize \nodata & \scriptsize \nodata & \scriptsize 1.96 & \scriptsize \nodata & \scriptsize Source confusion\\
\enddata
\tablecomments{
The upper limits in the ground-based images,
$UBVRIizy$, are defined by the $2\sigma$ level.
The numbers in parenthesis are
$3\sigma$ upper limits of \\
HST/ACS
$B_{435}$, $V_{606}$, $i_{775}$, and $z_{850}$ bands.
}
\tablenotetext{a}{
Our candidates falling 
in the central $\sim 160$ arcmin$^2$ field of GOODS-N
with HST and Spitzer images.
There are no obvious counterparts in IRAC $5.8\mu$m, \\
$8.0\mu$m, and
MIPS $24\mu$m bands (see Figure \ref{fig:image_zdrop_goodsn_HSTSpitzer}).
The $3\sigma$ upper limits in these bands
are $(m_{5.8},m_{8.0},m_{24})=(23.5,23.4,21.2)$.
}
\tablenotetext{b}{
Fluxes of these objects are contaminated by close bright objects on the sky.
}
\end{deluxetable}

\clearpage

\begin{deluxetable}{ccccccc}
\tablecolumns{7}
\tabletypesize{\scriptsize}
\tablecaption{UV Luminosity Function at $z=7$
\label{tab:muv_lumifun_schechter}}
\tablewidth{0pt}
\setlength{\tabcolsep}{0.02in}
\tablehead{
\colhead{$\phi^*$} &
\colhead{$M_{\rm UV}^*$} &
\colhead{$\alpha$} &
\colhead{Mag. Range} &
\colhead{$n^{\rm obs}$} &
\colhead{$\rho_{\rm UV}^{\rm obs}$} &
\colhead{$\rho_{\rm UV}^{\rm upper}$} \\
\colhead{($10^{-3}$Mpc$^{-3}$)} &
\colhead{(mag)} &
\colhead{} &
\colhead{(mag)} &
\colhead{($10^{-3}$Mpc$^{-3}$)} &
\colhead{($10^{25}$erg s$^{-1}$ Hz$^{-1}$Mpc$^{-3}$)} &
\colhead{($10^{25}$erg s$^{-1}$ Hz$^{-1}$Mpc$^{-3}$)} \\
\colhead{(1)} &
\colhead{(2)} &
\colhead{(3)} &
\colhead{(4)} &
\colhead{(5)} &
\colhead{(6)} &
\colhead{(7)} 
}
\startdata
$0.69_{-0.55}^{+2.62}$  & $-20.10\pm 0.76$  & $-1.72\pm 0.65$ & $-22.3<M<-18.0 $ & $2.2_{-1.3}^{+2.3}$ & $4.0_{-2.6}^{+6.2}$ & $10.6_{-5.3}^{+10.8}$ \\
\enddata
\tablecomments{
(1)-(3): Best-fit Schechter parameters.
The values of $\phi^*$ and $M^*_{\rm UV}$ are given
in units of $10^{-3}$Mpc$^{-3}$ and AB magnitude,
respectively. 
The reduced $\chi^2$ of the fitting is 0.13.
(4): Magnitude range of UV LFs that are used for the fitting. 
(5)-(6): Number density (in $10^{-3}$Mpc$^{-3}$)
and UV luminosity density (in $10^{25}$erg s$^{-1}$ Hz$^{-1}$ Mpc$^{-3}$)
calculated with the best-fit Schechter parameters down to
the limit of UV magnitude, 
$M_{\rm UV}\le -18$, defined by the HST/WFC3 studies.
(7): The upper limit UV luminosity density 
which is the integral of the best-fit Schechter function 
down to $M_{\rm UV}=\infty$. 
}
\end{deluxetable}

\end{document}